\documentclass[a4paper, oneside, 12pt]{scrartcl}

\usepackage[utf8x]{inputenc}
\usepackage[T1]{fontenc}
\usepackage[english]{babel}

\makeatletter
\DeclareOldFontCommand{\rm}{\normalfont\rmfamily}{\mathrm}
\DeclareOldFontCommand{\sf}{\normalfont\sffamily}{\mathsf}
\DeclareOldFontCommand{\tt}{\normalfont\ttfamily}{\mathtt}
\DeclareOldFontCommand{\bf}{\normalfont\bfseries}{\mathbf}
\DeclareOldFontCommand{\it}{\normalfont\itshape}{\mathit}
\DeclareOldFontCommand{\sl}{\normalfont\slshape}{\@nomath\sl}
\DeclareOldFontCommand{\sc}{\normalfont\scshape}{\@nomath\sc}
\makeatother

\usepackage[a4paper,  centering, bindingoffset=0mm, inner=25mm, outer=25mm, top=26mm, bottom=36mm, heightrounded]{geometry}

\usepackage{graphicx}
\usepackage{float}
\usepackage{cite}

\usepackage{amsmath}	
\usepackage{amssymb}
\usepackage{amsfonts}
\usepackage{mathrsfs}
\usepackage{mathtools}
\usepackage{amsthm}
\usepackage{bm}
\usepackage{esint}     

\usepackage{braket}
\usepackage{slashed}
\newcommand{\D}{{\cal D}}

\usepackage[font=small,labelfont=bf, width=.95\textwidth]{caption} 

\def\phg{
\begin{tikzpicture}[thick]
  \path [very thick, draw,snake it]
    (0,0)node[left]{$1^{\alpha}$} --(1,1);
  \draw[double, draw, snake it] (1,1)--(2.5,1)node[right]{$ 3^{\mu\nu}$};
  \path [very thick, draw,snake it]
    (1,1) --(0,2)node[left]{$2^{\beta}$};
\end{tikzpicture}
}
\def \loeik{
\begin{tikzpicture}[thick]
  \path [dashed, draw,snake it]
    (-2,0) --(0,0) -- (2,0);
  \draw[double, draw, snake it] (0,0)--(0,-2);
   \draw[->,,->=stealth, draw] (-0.3,-1.4)--(-.3,-.6) node[midway, left]
{$q$};
    \path [dotted, draw]
    (-2,-2) --(0,-2) -- (2,-2);
     \coordinate (A) at (0,-2);   \filldraw (A) circle (1.5pt);
\coordinate (B) at (0,0);   \filldraw (B) circle (1.5pt);
\end{tikzpicture}
}
\def \nloeikone {
\begin{tikzpicture}[thick]
  \path [dashed, draw,snake it]
    (-2,0) --(0,0) -- (2,0);
  \draw[double, draw, snake it] (0,0)--(0,-1) ;
  \draw[<-,,<-=stealth, draw] (-0.3,-.2)--(-0.3,-.8) node[midway, left]
{$q$};
  \draw[double, draw, snake it] (0,-1)--(-1,-2);
   \draw[<-,,<-=stealth, draw] (-0.6,-1.2)--(-1,-1.6) node[ above left]
{$k$};
  \draw[double, draw, snake it] (0,-1)--(1,-2);
  \draw[->,,->=stealth, draw] (0.6,-1.2)--(1,-1.6) node[ above right]
{$k-q$};
    \path [dashed, draw]    (-2,-2) --(0,-2) -- (2,-2);
 \coordinate (A) at (0,-1);   \filldraw (A) circle (1.5pt);
\coordinate (B) at (0,0);   \filldraw (B) circle (1.5pt);
\coordinate (C) at (-1,-2);   \filldraw (C) circle (1.5pt);
\coordinate (D) at (1,-2);   \filldraw (D) circle (1.5pt);
\end{tikzpicture}
}
\def \nloeiktwo{
\begin{tikzpicture}[thick]
  \path [dashed, draw,snake it]    (-2,0)--(0,0) -- (2,0);
  \draw[double, draw, snake it] (0,-1)--(0,-2);
   \draw[->,,->=stealth, draw] (1,-.4)--(.7,-.8) node[ midway,right]
{$k-q$};
  \draw[double, draw, snake it] (0,-1)--(1,0);
  \draw[<-,<-=stealth, draw] (-1,-.4)--(-.7,-.8) node[ midway,left]
{$k$};
  \draw[double, draw, snake it] (0,-1)--(-1,0);
   \draw[->,->=stealth, draw] (-0.3,-1.7)--(-0.3,-1.2) node[ midway,left]
{$q$};
    \path [dashed, draw] (-2,-2) --(0,-2) -- (2,-2);
  \coordinate (A) at (0,-1);
   \filldraw (A) circle (1.5pt);
\coordinate (B) at (1,0);
   \filldraw (B) circle (1.5pt);
\coordinate (C) at (-1,0);
   \filldraw (C) circle (1.5pt);
   \coordinate (D) at (0,-2);
   \filldraw (D) circle (1.5pt);
\end{tikzpicture}
}
\def\nloeikthree{\begin{tikzpicture}[thick]
  \path [dashed, draw,snake it]    (-2,0)--(0,0) -- (2,0);
  \draw[double, draw, snake it] (-1,0)--(-1,-2);
    \draw[<-,,<-=stealth, draw] (-1.3,-.6)--(-1.3,-1.2) node[midway, left]
{$k$};
  \draw[double, draw, snake it] (1,-2)--(1,0);
   \draw[->,,->=stealth, draw] (1.3,-.6)--(1.3,-1.2) node[midway, right]
{$k-q$};
  \draw[draw] (-1,-2)--(1,-2);
    \path [dashed, draw] (-2,-2) --(0,-2) -- (2,-2);
  \coordinate (A) at (1,-2);
   \filldraw (A) circle (1.5pt);
\coordinate (B) at (1,0);
   \filldraw (B) circle (1.5pt);
\coordinate (C) at (-1,0);
   \filldraw (C) circle (1.5pt);
\coordinate (D) at (-1,-2);
   \filldraw (D) circle (1.5pt);
\end{tikzpicture}
}
\def \nloeikfour{
\begin{tikzpicture}[thick]
  \path [dashed, draw,snake it]    (-2,0)--(-1,0);
   \path [dashed, draw,snake it]   (1,0)-- (2,0);
  \draw[double, draw, snake it] (-1,0)--(-1,-2);
   \draw[<-,,<-=stealth, draw] (-1.3,-.6)--(-1.3,-1.2) node[midway, left]
{$k$};
  \draw[double, draw, snake it] (1,-2)--(1,0);
   \draw[->,,->=stealth, draw] (1.3,-.6)--(1.3,-1.2) node[midway, right]
{$k-q$};
  \draw[draw,very thick, snake it] (-1,0)--(1,0);
  \path [dashed, draw] (-2,-2) --(0,-2) -- (2,-2);
  \coordinate (A) at (1,-2);
   \filldraw (A) circle (1.5pt);
\coordinate (B) at (1,0);
   \filldraw (B) circle (1.5pt);
\coordinate (C) at (-1,0);
   \filldraw (C) circle (1.5pt);
\coordinate (D) at (-1,-2);
   \filldraw (D) circle (1.5pt);
\end{tikzpicture}
}
\def \loimp{
\begin{tikzpicture}[thick]
  \path [dashed, draw,snake it]
    (-2,0) --(0,0);
    \path [very thick,draw,snake it]
    (0,0)--(2,0);
  \draw[double, draw, snake it] (0,0)--(0,-2);
   \draw[->,,->=stealth, draw] (-0.3,-1.4)--(-.3,-.6) node[midway, left]
{$q$};
    \path [dotted, draw]
    (-2,-2) --(0,-2) -- (2,-2);
\end{tikzpicture}
}
\def \nloimpone {
\begin{tikzpicture}[thick]
  \path [dashed, draw,snake it]
    (-2,0) --(0,0) ;
    \path [ very thick,draw,snake it]
    (0,0) -- (2,0);
  \draw[double, ,draw, snake it] (0,0)--(0,-1) ;
  \draw[<-,,<-=stealth, draw] (-0.3,-.2)--(-0.3,-.8) node[midway, left]
{$q$};
  \draw[double, draw, snake it] (0,-1)--(-1,-2);
   \draw[<-,,<-=stealth, draw] (-0.6,-1.2)--(-1,-1.6) node[ above left]
{$k$};
  \draw[double, draw, snake it] (0,-1)--(1,-2);
  \draw[->,,->=stealth, draw] (0.6,-1.2)--(1,-1.6) node[ above right]
{$k-q$};
    \path [dashed, draw]    (-2,-2) --(0,-2) -- (2,-2);
 \coordinate (A) at (0,-1);   \filldraw (A) circle (1.5pt);
\coordinate (B) at (0,0);   \filldraw (B) circle (1.5pt);
\coordinate (C) at (-1,-2);   \filldraw (C) circle (1.5pt);
\coordinate (D) at (1,-2);   \filldraw (D) circle (1.5pt);
\end{tikzpicture}
}
\def \nloimpfour{
\begin{tikzpicture}[thick]
  \path [dashed, draw,snake it]    (-2,0)--(-1,0);
   \path [very thick, draw,snake it]   (1,0)-- (2,0);
  \draw[double, draw, snake it] (-1,0)--(-1,-2);
   \draw[<-,,<-=stealth, draw] (-1.3,-.6)--(-1.3,-1.2) node[midway, left]
{$k$};
  \draw[double, draw, snake it] (1,-2)--(1,0);
   \draw[->,,->=stealth, draw] (1.3,-.6)--(1.3,-1.2) node[midway, right]
{$k-q$};
  \draw[draw,very thick, snake it] (-1,0)--(1,0);
  \path [dashed, draw] (-2,-2) --(0,-2) -- (2,-2);
  \coordinate (A) at (1,-2);
   \filldraw (A) circle (1.5pt);
\coordinate (B) at (1,0);
   \filldraw (B) circle (1.5pt);
\coordinate (C) at (-1,0);
   \filldraw (C) circle (1.5pt);
\coordinate (D) at (-1,-2);
   \filldraw (D) circle (1.5pt);
\end{tikzpicture}
}
\usepackage{verbatim}
\usepackage{textcomp}
\usepackage{enumitem}

\usepackage[toc]{appendix}

\usepackage{layout}
\usepackage{microtype}
\usepackage{bookmark}
\usepackage{color}
\usepackage[]{xcolor}

\newcommand{\dd}{\mathrm{d}} 
\newcommand{\ii}{\mathrm{i}}

\makeatletter
\newcommand{\subalign}[1]{%
  \vcenter{%
    \Let@ \restore@math@cr \default@tag
    \baselineskip\fontdimen10 \scriptfont\tw@
    \advance\baselineskip\fontdimen12 \scriptfont\tw@
    \lineskip\thr@@\fontdimen8 \scriptfont\thr@@
    \lineskiplimit\lineskip
    \ialign{\hfil$\m@th\scriptstyle##$&$\m@th\scriptstyle{}##$\crcr
      #1\crcr
    }%
  }
}
\makeatother
\usepackage{tikz}
\usetikzlibrary{decorations.pathmorphing}
\usetikzlibrary{shapes.geometric}
\usetikzlibrary{matrix, calc, arrows}
\tikzset{snake it/.style={decorate, decoration=snake}}
\usetikzlibrary{decorations.markings}

\tikzset{%
  dots/.style args={#1per #2}{%
    line cap=round,
    dash pattern=on 0 off #2/#1
  }
}

\definecolor{unamblue}{rgb}{0.0, 0.0, 0.0}

\usepackage[]{hyperref}
\hypersetup{
    colorlinks=true,%
    citecolor=unamblue,%
    filecolor=unamblue,%
    linkcolor=unamblue,%
    urlcolor=unamblue,
    bookmarksnumbered=true,     
    bookmarksopen=true,         
    bookmarksopenlevel=1,       
    pdfstartview=Fit,           
    pdfpagemode=UseOutlines,
    pdfpagelayout=TwoPageRight
}

\usepackage{graphicx}
\usepackage{tikz}
\usepackage{slashed}
\usepackage{epstopdf}
\usepackage{verbatim}	
\usepackage{blkarray}
\usepackage{ytableau}

\usepackage[]{todonotes}
\usepackage{subcaption}

\newcommand{\ba}{\begin{equation} \begin{aligned}}
\newcommand{\ea}{\end{aligned} \end{equation}}

\newcommand{\be}{\begin{equation}}
\newcommand{\ee}{\end{equation}}

\newcommand{\la}{\langle}
\newcommand{\ra}{\rangle}
\newcommand{\La}{\Big \langle}
\newcommand{\Ra}{\Big \rangle}

\newcommand{\deltahat}{	\hat{\delta}	}
\newcommand{ \bu} {\bar{u}}

\newcommand{\del}{\Delta}

\newcommand{\tree}{\begin{tikzpicture}[thick, scale=0.2]
\path [draw]
  (-2,1)--(-1,0) -- (1,0)--(2,1);
\path [draw]
  (-2,-1)--(-1,0)-- (1,0)--(2,-1);
\end{tikzpicture}
	}
\newcommand{\simtree}{{\scalebox{0.5}{\tree}}} 

\title{\Huge  Light bending from eikonal in worldline quantum field theory
\\}

\author{ \normalfont\normalsize Fiorenzo Bastianelli, Francesco Comberiati, Leonardo de la Cruz\\[2mm]
\emph{\normalfont\small \em Dipartimento di Fisica e Astronomia ``Augusto Righi'', Universit\`a di Bologna}\\
\emph{\normalfont\small \em and INFN Sezione di Bologna, via Irnerio 46, I-40126 Bologna, Italy}	
 }
\date{%
 $\,$%
    \\[2\baselineskip]
    \normalfont\normalsize%
      \parbox{0.8\linewidth}{%
{\bf \sf Abstract}. 
Using the worldline quantum field theory (WQFT) formalism for classical scattering, we study the deflection of light
by a heavy massive spinless/spinning object. WQFT requires the use of the worldline dressed propagator
of a photon in a gravitational background, which we construct from first principles.
The action required to set up the worldline path integral is constructed using auxiliary variables, 
which describe dynamically the spin degrees of freedom of the photon and take care of path ordering.
We test the fully regulated path integral  by recovering the photon--photon-graviton vertex.  
With the dressed propagator at hand, we follow the WQFT procedure by setting up the
partition function and deriving the  Feynman rules which can be used to evaluate it perturbatively.
These rules depend on the auxiliary variables. The latter ultimately do not contribute in the geometric-optics regime, which
realizes the equivalence between the scattering of a photon and  a massive scalar with that of a massless and a 
massive scalar. Then, the calculation of the eikonal phase and the 
deflection angle  simplifies considerably. Using the eikonal phase defined in terms of the partition function, we
calculate explicitly the deflection angle at NLO in the spinless case, and at LO in the spinning case up to quadratic order in spin.
}
}

\begin{document}
\maketitle
\thispagestyle{empty}
\newpage 
\tableofcontents
\section{Introduction}
The worldline quantum field theory (WQFT) description of classical scattering is a perturbative path integral formalism  
which simplifies the classical limit procedure of scattering amplitudes in gravity \cite{Mogull:2020sak}.
WQFT is based on a relation  between elastic (or inelastic) scattering amplitudes in the absence of matter loops and a worldline  path integral representation of the dressed Feynman propagator.\footnote{Roughly speaking, in perturbation theory, 
	a dressed propagator represents a resummation of tree-level Feynman diagrams of
	a particle propagating in a background (see Fig.~\ref{dress-prop}). Dressed propagators have been developed
	in a  worldline representation for a variety of models, see e.g. \cite{Daikouji:1995dz, Ahmadiniaz:2015kfq,
		Ahmadiniaz:2015xoa, Edwards:2017bte, Ahmadiniaz:2017rrk, Ahmadiniaz:2020wlm, Corradini:2020prz,
		Ahmadiniaz:2021gsd}.} 
The relation between the $S$-matrix, and dressed propagators requires a procedure to put  the latter on-shell 
after having removed the external legs. The prescription for obtaining such propagator was pioneered by Fradkin long time  ago~\cite{Fradkin:1966zz} and applied in Ref.~\cite{Fabbrichesi:1993kz} to study high energy scattering in gravity. 
Once the worldline  path integral is under control and the correspondence to the $S$-matrix made explicit, expectation values can be computed from a partition function expressed as a worldline path integral. One can then derive Feynman rules of the theory which allow the calculation of these expectation values directly. This is  the WQFT approach to classical scattering observables. WQFT shares some similarities with the   Effective Field Theory (EFT) approach to gravitational dynamics \cite{Goldberger:2004jt,
	Goldberger:2006bd, Goldberger:2009qd} with the important difference that in WQFT worldline degrees of freedom are also 
quantized. The WQFT formalism  has been further developed to describe Bremsstrahlung \cite{Jakobsen:2021smu}, spinning black holes \cite{Jakobsen:2021lvp, Jakobsen:2021zvh} and colored massive particles \cite{Shi:2021qsb}. 

The direct relation between the WQFT formalism  and on-shell scattering amplitudes motivate us to find  
a similar treatment of the classical scattering of light in the presence of a scalar/spinning object\footnote{The  different approaches to describe this system and their relation with the eikonal phase is nicely reviewed in Ref.~\cite{Bjerrum-Bohr:2017dxw}.}. Treating quantum gravity as an EFT \cite{Donoghue:1994dn,Bjerrum-Bohr:2002gqz,Bjerrum-Bohr:2013bxa} it was  recently found that there are  small quantum  corrections that differentiate the scattering of a massless particle from  that of light off a heavy mass object \cite{Bjerrum-Bohr:2014zsa, Bjerrum-Bohr:2016hpa, Bai:2016ivl}, thus testing the equivalence principle \cite{Bjerrum-Bohr:2015vda}. Since these differences are absent in the classical limit, one can recover  deflection angles by taking the massless limit of the  scattering of massive particles, 
see e.g. Ref.\cite{Bjerrum-Bohr:2018xdl}. In particular, as studied a long time ago by Amati, Ciafaloni and Veneziano \cite{Amati:1990xe},
one can employ the eikonal phase  to extract classical deflection angles in gravity through differentiation.
More recently, this approach has received a renewed interest \cite{Melville:2013qca,Luna:2016idw,Akhoury:2013yua,KoemansCollado:2019ggb, Cristofoli:2020uzm,DiVecchia:2019myk,DiVecchia:2019kta,Bern:2020gjj,Parra-Martinez:2020dzs,DiVecchia:2021bdo,Heissenberg:2021tzo,Damgaard:2021ipf, delaCruz:2021gjp,Bern:2020buy, AccettulliHuber:2020oou, Kosmopoulos:2021zoq,Brandhuber:2021eyq, Aoude:2021oqj}. Another route to the problem was taken in Ref.~\cite{Cristofoli:2021vyo}, where  
the Kosower-Maybee-O'Connell (KMOC) formalism \cite{Kosower:2018adc, Maybee:2019jus, delaCruz:2020bbn} for classical observables
was generalized to describe the classical limit of massless particles. In one of the applications of
Ref.~\cite{Cristofoli:2021vyo}, 
the geometric-optics regime was used to  extract the deflection angle from
scattering amplitudes,  since it is in this  limit in which the amplitude can be related to a beam of light through the
precise definition of the classical observable. 
As we shall see, our WQFT construction requires this exact  regime to extract the deflection angle through the eikonal phase.

In this paper, we construct the appropriate WQFT to deal with photons  with a first goal of
realizing the equivalence between the scattering of a photon and  a massive scalar with that of a massless and a 
massive scalar in the geometric-optics regime.
The second goal is a simplified approach to the calculation of the scattering angle. 
Unlike  the case of a matter particle propagating in a gravitational background,  the  dressed photon propagator depends on a matrix-valued action and therefore the worldline path integral must include a path ordering. The path ordering can be avoided by rewriting the  path integral in terms of auxiliary variables at the cost of introducing additional integrals. 
These variables  are inherently quantum---  they describe the quantum polarization of the
spin 1 particle--- but ultimately they do not play any role in our classical computation. This insight plays a key role to show the aforementioned equivalence. 

Once the dressed propagator is obtained and put on-shell, we can follow the  WQFT setup to derive its Feynman rules and compute the eikonal phase.
The Feynman rules that we find depend in general on the auxiliary variables.
Building on the insights of Ref.\cite{Cristofoli:2021vyo},  we  consider the geometric-optics regime, which 
leads to a great simplification of the calculations.  As we shall see, this regime implies  
the vanishing of terms proportional to the spin-tensor and serves as a check of our setup.
Then, the calculation of the eikonal phase and derived quantities, such as the deflection angle and the  impulse, dramatically simplifies.

The rest of the paper  is organized as follows: In section \ref{review}, we review the WQFT formalism and introduce our conventions.   In Section \ref{dressed-propagator}, we derive the dressed photon propagator required for the 
implementation of WQFT and define the eikonal phase. 
In Section \ref{calculations} we compute the eikonal phase and deflection angles 
at LO and NLO. Our conclusions are presented in Section 
\ref{conclusions}.

\section{Review}
\label{review}
\subsection{The  gravitationally dressed scalar propagator} \label{2.1}
In order to introduce our notation and conventions, let us consider first the case of a single scalar massive particle of mass $m$. We will use the mostly minus signature for the Minkowski metric $\eta_{\mu\nu}=\text{diag}(1, -1,-1,-1)$
and set the gravitational coupling to  $\kappa^2=32\pi G_N$, where $G_N$ is the Newton constant. The gravitational action is given by the usual Einstein-Hilbert action 
\begin{align}
S_{\text{EH}}=- \frac{2}{ \kappa^2} \int \dd^4 x \sqrt{-g}R \;,
\end{align}
whereas the action for the massive scalar field including a non-minimal coupling of the scalar field to the background curvature is given by
\begin{align}
S_{\text{m}}=\int \dd^4 x \sqrt{-g} \left[ g^{\mu\nu}\partial_\mu \varphi^*\partial_\nu\varphi
+ (\xi R - m^2) \varphi^{*}\varphi \right].
\label{KG-action}
\end{align}
Here  $\xi$ is a free dimensionless coupling. 
Requiring Weyl invariance in the massless case fixes this coupling  
to $\xi=\frac16$ ($\xi=\frac{d-2}{4(d-1)}$ in arbitrary dimensions),
but here we shall keep it arbitrary.
In order to relate scattering amplitudes and path integrals, 
we first rewrite the scalar propagator in an external gravitational field 
in a proper time representation 
\begin{align}
\ii  G(x, y;g)= \braket{y| \frac{1}{ \hat H-\ii \epsilon }|x}= 
\ii \int_{0}^{\infty} \dd T \braket{y| e^{-\ii T( \hat H -\ii \epsilon)} |x},
\label{FS-proper-time}
\end{align}
where $T$ is the Fock-Schwinger proper time. The Hamiltonian operator $\hat H$
corresponds to the Klein-Gordon operator fixed by the action \eqref{KG-action}
and is given by 
\begin{align}
\hat H= g^{\mu\nu} \nabla_\mu  \nabla_\nu +m^2 -\xi R = \frac{1}{\sqrt{-g}}\partial_\mu \sqrt{-g} g^{\mu\nu}\partial_\nu+m^2 -\xi R \;.
\end{align}
It can be viewed as arising from  a classical particle Hamiltonian obtained by setting $\partial_\mu \to - \ii p_\mu$ in the last expression, 
finding $H=-g^{\mu\nu}p_{\mu}p_{\nu}+m^2-\xi R$. The particle action in hamiltonian form can be written as 
\begin{align}
S_{\text {p}}= \int_0^1 \dd \tau (p_\mu \dot x^\mu -e H),
\label{worldline-action-1}
\end{align}
where $e$ is the einbein  that gauges translations on the worldline, leading to a reparametrization invariant description of the worldline. The einbein reproduces the effect of the proper time $T$ upon gauge fixing $e(\tau)=T$. Then, rescaling the proper time $\tau$ to range in the interval $[0, T]$, we obtain
the particle action in configuration space 
\begin{align}
S_{\text{p}}= \int_{0}^T \dd \tau \left[-\frac{1}{4}g_{\mu\nu} \dot 
x^\mu \dot x^\nu- m^2 +\xi R \right] \;.
\label{worldline-action-n1}
\end{align}

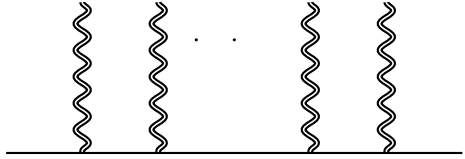
\begin{figure}
	\centering
	\begin{tikzpicture}[thick]
	\draw[double, draw, snake it] (-2,0)--(-2,-2);
	\draw[double, draw, snake it] (-1,0)--(-1,-2);
	\draw[double, draw, snake it] (2,0)--(2,-2);
	\draw[double, draw, snake it] (1,0)--(1,-2);
	\path [draw]
	(-3,-2) --(0,-2) -- (3,-2);
	\path [dashed, very thick, dots=2 per 1cm, draw](-0.5,-0.5) -- (0.5,-0.5);
	\end{tikzpicture}
	\caption{Dressed propagator with external massive particles off-shell}
	\label{dress-prop}
\end{figure}

In order to define a path integral free of spurious UV divergences and regularization ambiguities, one must introduce auxiliary worldline ghost variables and a finite counterterm to the worldline action \eqref{worldline-action-n1}. Let us mention that these issues are only related to the one-dimensional worldline theory, and are not  related to regularization of spacetime. The case of UV divergences on the worldline can be addressed by defining the path integral measure as follows 
\begin{equation}
\mathcal{D} x := D x \prod\limits_{0 <\tau  < T} \sqrt{- g(x(\tau))}=
D x \int D a D b Dc\ \exp\left[-\ii \int_0^T \dd\tau \frac{1}{4} g_{\mu\nu}(a^\mu a^\nu+ b^\mu c^\nu )\right], 
\end{equation}
where the final form contains the standard translationally invariant measures,
indicated by the symbol $D$ as opposed to the symbol  $\mathcal{D}$.
The second equality exponentiates the 
determinant factor and leads to standard perturbation theory on the worldline. 
As regularization ambiguities  play no role in the upcoming discussion, let us just  mention that three options to fix such ambiguities are  known,
and correspond to the
time slicing (TS) regularization, mode regularization (MR), and worldline dimensional regularization (DR)  \cite{Bastianelli:2006rx}.
The appropriate counterterms in these schemes can be written as follows
\begin{align}
S_{\text{CT}}=\int_0^T  \dd \tau \left(-\frac{1}{4}R-V_{\text{TS/MR/DR}}\right),
\label{ct}
\end{align}
where the additional terms $V_{\text{TS/MR/DR}}$ are scheme dependent\footnote{They are
	given by $V_{\text{TS}}=-1/4\, g^{\mu\nu} \Gamma^{\beta}_{\mu\alpha}\Gamma^{\alpha}_{\nu\beta}$, 
	$V_{\text{MR}}=1/12\, g^{\mu\nu}g^{\alpha\beta } g_{\rho\sigma} \Gamma^{\rho}_{\mu\alpha}
	\Gamma^{\sigma}_{\nu\beta}$, and $V_{\text{DR}}=0$.
	Counterterms for supersymmetric versions of the nonlinear sigma model can be found in \cite{Bastianelli:2011cc}.}. 
The path integral in configuration space associated with the propagator is finally given by (see Fig.\ref{dress-prop}).
\begin{align} \label{prop}
G(x, y;g)=& \int_{0}^{\infty} \dd T e^{-\ii m^2 T} \int_{x(0)=x}^{x(T)=y} D x \int D a D b 
Dc \\
&  \exp \left\{-\ii \int_{0}^T \dd \tau \left[\frac{1}{4}\left(g_{\mu\nu} \dot 
x^\mu \dot x^\nu+a^\mu a^\nu+ b^\mu c^\nu \right)+\left(\frac{1}{4} - \xi\right) R+V_{\text{TS/MR/DR}}\right]\right\} \;.
\nonumber
\end{align}
It can be solved  in perturbation theory with standard gaussian integration 
\cite{Bastianelli:2002fv, Schubert:2001he}.  
In order to apply this technique,  we consider the  splitting of the position variable into a straight 
line  and a quantum fluctuation $q^\mu$
\begin{equation}
x^\mu(\tau) =b^\mu+ v^\mu \tau + q^\mu(\tau) .\;
\end{equation}
At this point the dressed propagator represents a resummation of tree-level Feynman diagrams in which the 
straight line parameters $b^\mu$ and $u^\mu$ depend on boundary conditions one wishes to impose,
and the quantum-fluctuation variables acquire Dirichlet boundary conditions (DBC), namely $q(0)=q(T)=0$. 
As we will review shortly, relating dressed propagators and 
scattering amplitudes in the classical limit requires to appropriately express 
the boundary conditions carried by the parameters of the straight line in terms of the 
physical parameters of the scattering.

\subsection{Relation to scattering amplitudes}
\label{review-on-shell}
Consider the  elastic scattering of two spinless massive particles of masses $m_1$ and $m_2$ in the absence of matter loops. Each massive particle is described by  his own dressed propagator, which we can write as $G_i(x, y;g)$, $i=1,2$ with obvious labeling of 
the worldline variables and masses. The dressed propagators are connected but not amputated. In order to define 
the $S$-matrix we must amputate the external legs and put them on-shell. In momentum space the amputated on-shell dressed propagator  reads 
\begin{align}
G^{\text{c}}(p,p';g):= \lim\limits_{p^2, p'^2 \to m^2} \ii(p^2-m^2) \ \ii ({p'}^2-m^2) \int \dd^4 x \dd^4y \
e^{\ii p\cdot x-\ii p'\cdot y}G(x, y;g) \;.
\end{align}
The overall effect of putting the dressed propagator on-shell amounts to 
eliminate the proper time integral in \eqref{prop} and set the region of integration over  $\tau$
in the action to $\tau \in (-\infty, +\infty)$, as shown in Ref.~\cite{Mogull:2020sak}. 

\begin{figure}
	\centering
	\begin{subfigure}{0.4\textwidth}
		\centering
		\begin{tikzpicture}[thick]
		\path [draw]
		(-2,0) --(0,0) -- (2,0);
		\node[ellipse, draw, fill = gray!20] (e) at (0,-1) {Trees};
		\draw[double, draw, snake it] (-1,0)--(e.north west);
		\draw[double, draw, snake it] (0,0)--(e.north);
		\draw[double, draw, snake it] (1,0)--(e.north east);
		\draw[double, draw, snake it] (-1,-2)--(e.south west);
		\draw[double, draw, snake it] (0,-2)--(e.south);
		\draw[double, draw, snake it] (1,-2)--(e.south east);
		\path [ draw](-2,-2) --(0,-2) -- (2,-2);
		\path [dashed, very thick, dots=4 per 1cm, draw](-0.6,-0.5) -- (0,-0.5);
		\path [dashed, very thick, dots=4 per 1cm, draw](-0.6,-1.6) --
		(0,-1.6);
		\end{tikzpicture}
		\caption{elastic}
	\end{subfigure}
	\begin{subfigure}{0.4\textwidth}
		\centering
		\begin{tikzpicture}[thick]
		\path [draw]
		(-2,0) --(0,0) -- (2,0);
		\node[ellipse, draw, fill = gray!20] (e) at (0,-1) {Trees};
		\draw[double, draw, snake it] (-1,0)--(e.north west);
		\draw[double, draw, snake it] (0,0)--(e.north);
		\draw[double, draw, snake it] (1,0)--(e.north east);
		\draw[double, draw, snake it] (-1,-2)--(e.south west);
		\draw[double, draw, snake it] (0,-2)--(e.south);
		\draw[double, draw, snake it] (1,-2)--(e.south east);
		\draw[double, draw, snake it] (2,-1)--(e.0);
		\path [ draw](-2,-2) --(0,-2) -- (2,-2);
		\path [dashed, very thick, dots=4 per 1cm, draw](-0.6,-0.5) -- (0,-0.5);
		\path [dashed, very thick, dots=4 per 1cm, draw](-0.6,-1.6) --
		(0,-1.6);
		\end{tikzpicture}
		\caption{inelastic}
	\end{subfigure}
	\caption{Diagrams resummed by the method. The blob represent trees including disconnected ones. }
	\label{trees-diagrams}
\end{figure}
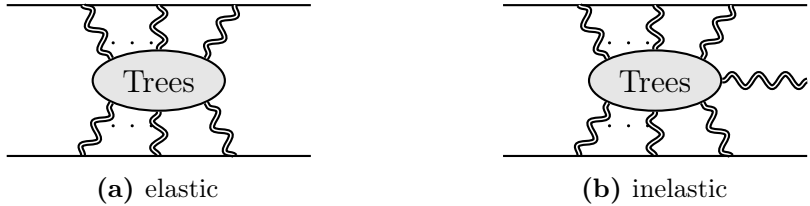

Let us now consider the  weak field approximation $g_{\mu\nu}=\eta_{\mu\nu}+\kappa h_{\mu\nu}$, and add to the 
Einstein-Hilbert action the gauge-fixing term 
\begin{align}
S_{\text{gf}}=\int \dd^4 x\left(\partial^{\nu}h_{\mu\nu}-1/2 \partial_\mu h^\nu_{\ \nu} \right)^2,
\end{align}
which imposes a weighted version of the de Donder gauge $\partial^{\nu}h_{\mu\nu}=1/2 \partial_\mu h^\nu_{\ \nu}$. The full action is then
\begin{align}
S_g=S_{\text{EH}}+S_{\text{gf}} \;.
\label{action-grav-gauge}
\end{align}
The on-shell version of the dressed propagators can  then be used to relate the elastic 
scattering amplitude in the absence of matter loops as follows
\begin{align}
\mathcal{A} (p_1 p_2 \to p_1'p_2')
=\mathcal{N} \int \D g_{\mu\nu} \ e^{\ii S_g} G^{c}_1(p_1, p_1';g) G^{c}_2(p_1, p_2';g) \;,
\label{amp-to-path}
\end{align}
where $\mathcal{N}$ is a normalization factor. The classical limit on the right hand side is taken and
understood as the usual one in  QFT, i.e., considering only graviton Born diagrams (See Fig.~\ref{trees-diagrams}).  To avoid the proliferation of factors of $2\pi$, we will adopt the short-hand notation 
\begin{align}
\hat \dd^n x := \frac{ \dd^n x}{(2\pi)^n}, \qquad  \hat \delta^n(x):= (2\pi)^n \delta^n (x) \;.
\end{align}

We define the eikonal phase as in Refs.~\cite{Amati:1990xe, Amati:1987wq} by
\begin{align}
e^{\ii \chi} := \frac{1}{4m_1 m_2} \int \hat{\dd}^d q \  \hat{\delta}(q \cdot v_1) \hat{\delta}(q \cdot v_2) e^{\ii q \cdot b }\mathcal{A} (p_1 p_2 \to p_1'p_2') \;,
\end{align}
where $b:=b_2-b_1$ is the impact parameter and $q:=p_1'-p_1=p_2-p_2'$ is the momentum exchange. The  
eikonal phase and the QFT S-matrix (in the classical limit) in the worldline theory can be nicely 
related as \cite{Mogull:2020sak}
\begin{align}
\mathcal{Z}_{\text{WFT}}=e^{\ii \chi},
\end{align}
which links the free energy of the WQFT and the eikonal phase. The path integral representation of the partition function  can be written as 
\begin{align}
\mathcal{Z}_{\text{WFT}}:=& \mathcal{N} \int D h_{\mu\nu} \int  \prod\limits_{i=1}^2
D x_i \, \int 
\prod\limits_{i=1}^2 (D a_i D b_i D c_i ) \ e^{\ii S_{g}}
\label{partition-function-scalar}
\\
& \quad \exp\left[-\ii \sum\limits_{i=1}^2 \int^\infty_{-\infty}\dd \tau_i 
\frac{m_i}{2}g_{\mu \nu}(\dot x_i^\mu \dot x_i^\nu+ a_i^\mu a_i^\nu+b_i^\mu c_i^\nu)\right],\nonumber
\end{align}
where  the normalization factor $\mathcal{N}$ ensures that $\mathcal{Z}_{\text{WFT}}=1$ in the free theory. Expectation values of operators $\mathcal{O}(h, \{x_i\})$ are obtained as
\begin{align}
\big\langle \mathcal{O}(h, \{x_i\}) \big\rangle_{\text{WQFT}}=& 
\mathcal{Z}_{\text{WFT}}^{-1} \int D h_{\mu\nu} \int  \prod\limits_{i=1}^2
D x_i \, \int 
\prod\limits_{i=1}^2 (D a_i D b_i D c_i ) \ e^{\ii S_{g}}\mathcal{O}(h, \{x_i\})
\\
& \quad \exp\left[-\ii \sum\limits_{i=1}^2 \int^\infty_{-\infty}\dd \tau_i 
\frac{m_i}{2}g_{\mu \nu}(\dot x_i^\mu \dot x_i^\nu+ a_i^\mu a_i^\nu+b_i^\mu c_i^\nu)\right].\nonumber
\end{align}

\subsection{Worldline Feynman rules}
The path integral can be solved perturbatively using worldline Feynman rules which keep track of the 
expansion in power of the coupling $\kappa$ and take care of the Wick contractions between fields
and the quantum fluctuations $q_i$.  The role of the auxiliary ghost variables $a_i$, $b_i$, $c_i$ is to cancel 
divergences which are of quantum nature. Since we are not interested in those corrections, it is enough to 
summarize the Feynman rules associated with the fields $h_{\mu\nu}$ and the deflections $q^\mu(\tau)$.
In order to make contact more naturally with on-shell scattering amplitudes, we scale the worldline parameters of the particle action as $\tau_{i} \to \sigma_{i}/m_{i}$ such that the background split in configuration space reads    
\be
x_{i}^{\mu}(\sigma_{i}) = b_{i}^{\mu} +p_{i}^{\mu}\sigma_{i} + q_{i}^{\mu}(\sigma_{i}) \;,
\ee
where $p_{i}^{\mu}= m_{i} v_{i}^{\mu}$. The Fourier transforms to momentum space and energy space\footnote{Given the rescaling we performed in the worldline action, the worldline parameter $[\sigma]\sim L^{2}$ then implies that $[\omega]\sim L^{-2}$ to keep dimensionless the plane wave exponential in Fourier transforms.} for $h_{\mu\nu}(x)$ and $q^{\mu}(\sigma)$, respectively
are  defined by
\begin{align}
h_{\mu\nu}(x)=\int \hat{\dd}^4 \ell \ e^{-\ii \ell \cdot x} h_{\mu\nu}(\ell), \qquad 
q^\mu(\sigma)= \int \hat{\dd} \omega \ e^{-\ii \omega \sigma}q^\mu(\omega).
\end{align}

Let $P_{\mu\nu\rho\sigma} :=  (\eta_{\mu(\rho} \eta_{\nu)\sigma}- \frac{1}{2}\eta_{\mu\nu}\eta_{\rho\sigma})$, where
the parenthesis of Lorentz indices denotes symmetrization with unit weight, e.g.~$v_1^{(\mu} v_2^{\nu)} = \frac{1}{2}(v_1^{\mu} v_2^{\nu} + v_1^{\nu} v_2^{\mu})$. The propagators then read
\begin{align}
\label{eq:propq}\raisebox{-2mm}{\begin{tikzpicture}[thick]
	\coordinate (A) at (-1,-0);
	\coordinate (B) at (1,-0);
	\filldraw (A) circle (2pt) node[left] {$q^\mu$};
	\filldraw (B) circle (2pt) node[right] {$q^\nu$};
	\draw[->,>=stealth] (-0.5,0.2) -- (0.5,0.2) node[midway,above] {$\omega$};
	\draw (-1,0) -- (1,0);
	\end{tikzpicture}
}
= - \ii\frac{ \eta^{\mu \nu}}{2}\left( \frac{1}{(\omega+\ii\epsilon)^2}+\frac{1}{(\omega-\ii\epsilon)^2} \right), \
\raisebox{-2mm}{\begin{tikzpicture}[thick]
	\path [double, draw,snake it]
	(-1,0)node[left]{$h_{\mu\nu}$} --(0,0) -- (1,0)node[right]{$h_{\rho\sigma}$};
	\end{tikzpicture}
}     = \ii  \frac{P_{\mu\nu\rho\sigma}}{k^2+\ii \epsilon} , 
\end{align}
i.e., we use time-symmetric propagator for worldlines and Feynman propagators for gravitons. The choice
of $\ii \epsilon$ prescription determines the precise interpretation of the background parameters. Let us 
briefly summarize the findings of Ref.~\cite{Mogull:2020sak} (see also Ref.~\cite{Jakobsen:2021lvp}).
With  retarded (advanced) propagators the background parameters $b^\mu$, $v^\mu$ are associated with their undeflected trajectories
when $\tau \to -\infty$ ($\tau \to +\infty$). The time-symmetric prescription averages between these options 
\begin{align}
v^\mu=\frac{1}{2} \left( v^\mu_{-\infty}+v^\mu_{+\infty}\right)+\mathcal{O}(G_N^2), \qquad 
b^\mu=\frac{1}{2} \left( b^\mu_{-\infty}+b^\mu_{+\infty}\right)+\mathcal{O}(G_N^2),
\end{align}
which are the background parameters we will use in our calculations of  scattering angles from the 
eikonal phase. Therefore, the relation of the momenta of massive particles $p_i^\mu=m_i v_i$ is associated with 
the average four velocities.  At the maximum order we consider (2PM) in this paper, the difference between the far past or future background parameters are of order $\mathcal{O}(G_N)$ and it follows that  $b^\mu_{\pm\infty} = b^\mu+ \mathcal{O}(G_N)$. Therefore it will be not necessary to distinguish between $|b_{-\infty}|$ and 
$|b|$.  The interactions are then  described by the rules
\begin{align}
\label{eq:vertex0z1hDG}
\raisebox{-10mm}{\begin{tikzpicture}[thick]
	\path [dashed, draw]
	(-1,-1) -- (1,-1)node[right]
	{$\quad $};
	\path [double, draw,snake it]
	(0,-1) -- (0,-2)node[below]{$h_{\mu\nu}$};
	\draw[->,>=stealth] (-0.5,-1.2) -- (-0.5,-1.8) node[midway, left]
	{$k$};
	\coordinate (A) at (0,-1);
	\filldraw (A) circle (1.5pt);
	\end{tikzpicture}}&
= -\ii \frac{ \kappa}{2} e^{\ii k \cdot b} \hat \delta(k \cdot p) p^{\mu} p^{\nu},\\
\raisebox{-10mm}{\begin{tikzpicture}[thick]
	\path [dashed, draw]
	(-1,-1) -- (0,-1);
	\path [draw]
	(0,-1) -- (1,-1) node[right]
	{$q^\rho(\omega)$};
	\draw[->,>=stealth] (0.2,-.7) -- (.8,-.7) node[midway, above]
	{$\omega$};
	\path [double, draw,snake it]
	(0,-1) -- (0,-2)node[below]{$h_{\mu\nu}$};
	\draw[->,>=stealth] (-0.5,-1.2) -- (-0.5,-1.8) node[midway, left]
	{$k$};
	\coordinate (A) at (0,-1);
	\filldraw (A) circle (1.5pt);
	\end{tikzpicture}} &
\begin{aligned}
= \frac{\kappa}{2} e^{\ii k \cdot b} \hat \delta(k \cdot p+  \omega)  \left(2 
\omega p^{(\mu} \delta_{\rho}^{\nu)}+p^{\mu} p^{\nu} k_{\rho}\right).
\end{aligned}
\end{align}
Additional Feynman rules with more fluctuations along the worldline arise from higher order terms
of the perturbative expansion. They are required
for the calculation of the impulse and radiation. The interested reader can check Ref.~\cite{Mogull:2020sak}

\section{Photons and WQFT}
\label{dressed-propagator}
Our goal is to derive the partition function that describes scattering of photons and massive scalar/spinning 
particles in a gravitational background. To achieve this goal, we derive  the 
photon propagator dressed by gravitons and its worldline path integral 
representation. Then, we can use this expression to express the amplitude in terms of path integrals as in 
Eq.\eqref{amp-to-path} where one of the dressed propagators is associated with a photon. 

\subsection{Derivation of the gravitationally  dressed  photon propagator}
Let us now add to the Einstein-Hilbert action, the action of Maxwell theory minimally coupled to gravity
\be\label{EM-action}
S_{\gamma} =  -\frac{1}{4} \int \dd^{4}x \sqrt{-g} \,g^{\mu\alpha}g^{\nu\beta} F_{\mu\nu}F_{\alpha\beta}\;.
\ee
Its gauge symmetry, $ \delta A_{\mu}(x)  = \partial_{\mu}\alpha(x)$, $\delta g_{\mu\nu}(x) = 0$, 
can be covariantly gauge-fixed using standard BRST methods. The procedure is as follows.
One replaces the gauge parameter $\alpha(x)$ by the anticommuting ghost $c(x)$, and from the gauge algebra
obtains  the anticommuting  BRST variation $s$ (the so-called Slavnov variation).
It is required to be nilpotent, 
and is then extended to the non-minimal fields needed for gauge fixing, the antighost $\bar{c}(x)$  and auxiliary $B(x)$,
which are Gra{\ss}mann odd and even, respectively. The BRST symmetry is  
\be 
s A_\mu = \partial_\mu c \ , \ \
s c = 0 \ , \ \
s \bar{c} = B \ , \ \
s B=0, 
\ee
and can be easily verified to be nilpotent ($s^2 =0$). It is used to obtain the gauge-fixed total action
by adding to the lagrangian contained in \eqref{EM-action} 
the manifestly BRST invariant term  $s \Psi_\xi$,
where 
$\Psi_{\xi} = \sqrt{-g} \, \bar{c}\left(\nabla^{\mu}A_{\mu} +\frac{\xi}{2}B	\right)$
is the gauge fermion chosen to produce a $R_\xi$ gauge in curved space. The fields $c$, $\bar c$ and $B$ are all taken to be scalars under change of coordinates, 
so to keep covariance manifest. One finds 
\be
s \Psi_\xi =  \sqrt{-g} \left ( B \nabla^{\mu}A_{\mu}+\frac{\xi}{2}B^{2} - \bar{c}\nabla^{\mu}\partial_{\mu}c \right ) \sim 
\sqrt{-g} \left ( -\frac{1}{2\xi} (\nabla^{\mu}A_{\mu})^2 - \bar{c}\nabla^{\mu}\partial_{\mu}c \right ), 
\ee
where in the last step the auxiliary field $B$  has been eliminated by its own algebraic equations of motion.
We choose the value $\xi =1$, that implements the Feynman gauge. The total gauge-fixed BRST invariant action, $S_\text{tot}=S_\gamma+\int \dd^4 x \, s \Psi_\xi$, contains a ghost action that we disregard (at tree-level ghosts do not contribute) and terms that identify the photon propagator reads
\be
S_{\gamma,\, \text{gf}} =   
\int \dd^{4}x \sqrt{-g} \left [  -\frac{1}{4}  F_{\mu\nu} F^{\mu\nu}  - \frac12  (\nabla^{\mu}A_{\mu})^2 \right ]
=  \int \dd^{4}x \sqrt{-g} \left [  \frac12 A^{\mu}  \hat{H}_\mu{}^\nu A_\nu \right ],
\ee
where the second  form is obtained by performing partial integrations. It produces the second order differential operator 
\begin{align} 
\hat{H}_\mu{}^\nu = \delta_\mu{}^\nu\nabla^{2}-  R_\mu{}^\nu \;,
\end{align}
whose inverse gives the photon propagator in a curved background,
i.e. the dressed propagator of our interest.
The operator  $\hat{H}_\mu{}^\nu$
may be interpreted as a first-quantized Hamiltonian. 
For that purpose, we find it convenient to use flat indices by introducing a vielbein $e_\mu^a(x)$,
so the metric is given by $g_{\mu\nu}(x) = \eta_{ab} e^{a}_{\mu}(x) e^{b}_{\nu}(x)$.
This allows to present the Hamiltonian as  
\begin{align}
\hat{H}_a{}^b =\delta_a{}^b\nabla^2 - R_a{}^b,
\label{operator-ham}
\end{align}
where the  covariant derivative contains also the spin connection $\omega_{\mu a}{}^b(x)$,  
as it acts on vectors $A_a(x)= e^{\mu}_{a}(x)A_\mu(x)$ with flat indices, so the covariant derivative is $\nabla_\mu A_a = \partial_\mu A_a + \omega_{\mu a}{}^b A_b$.
On a more general Lorentz tensor, the covariant derivative takes the form $\nabla_\mu= \partial_\mu -\frac{\ii}{2} \omega_\mu{}^{ab} S_{ab}$, with $S_{ab}$
the Lorentz generators in the representation of the tensor.

Then, as in  Sec.~\ref{2.1}, we find that the Hamiltonian \eqref{operator-ham} can be reproduced by quantization of a  
relativistic particle now with matrix-valued action given by 
\begin{align}
(S_{\text{p}}[x;g])_a{}^b = \int_{0}^{1} \dd\tau \left(
-\frac{1}{4T}g_{\mu\nu}\dot{x}^{\mu}\dot{x}^{\nu} \delta_a{}^b -\frac{1}{2} 
\dot{x}^{\mu}\omega_{\mu}^{cd} (S_{cd})_a{}^b +T R_a{}^b -\frac{1}{4}TR\delta_a{}^b\right) \;,
\label{mva}
\end{align}
which contains the Lorentz generator  in the spin-1 representation
\begin{align}
(S_{cd})_{a}\,^{b} = \ii \left( \eta_{ca}\delta_{d}\,^{b} - \eta_{da}\delta_{c}\,^{b}\right) \;. 
\end{align}
We have already gauge-fixed the einbein $e(\tau)$, required to have a reparametrization invariant 
description of the worldline, to the Fock-Schwinger proper time $T$,
used in heat kernel methods to exponentiate the inverse
of a differential operator as in \eqref{FS-proper-time}.
The last term in the action is the counter-term for regularization (DR) of
the path integral, that we have already anticipated.
It is the same one that appears in the scalar particle case, see Eq.~\eqref{ct}.

The particle action leads to a quantum mechanical transition amplitude, represented by the following
worldline path integral
\be
D_{a}\,^{b}(x_{0},y_{0};g)=\int_{0}^{\infty} \dd T \braket{y_0, a| e^{-\ii T \hat{H}}|x_0,b } = \int_{0}^{\infty} \dd T \int_{x(0)=x_{0}}^{x(1)=y_{0}} \hskip -.2cm \D x \,  \mathrm{\bm T} \, \left( e^{\ii S_{\text{p}}[x;g]} \right)_{a}\,^{b},
\ee
where $\textrm{T}$ denotes path ordering. It furnishes a representation of the 
photon propagator in a graviton background. 
The path ordering prescription in the path integral generates the correct gauge transformation 
of the quantum mechanical transition amplitude, which behaves as a bi-tensor under the 
local Lorentz gauge group that acts of the flat indices
(see Ref.~\cite{Bastianelli:1992ct} for more details on this worldline model and the counterterm).

For perturbative calculations, we find it useful to introduce auxiliary bosonic 
variables $Q_a$ and $\bar Q^a$, that allow to treat dynamically the spin degrees of freedom 
of the photon while encoding simultaneously  the time-ordering 
prescription\footnote{The same construction can be done using fermionic variables as well.}.
The quantization of these variables produces  an enlarged Hilbert space,
which must be reduced to the appropriate one corresponding 
to the spin of the photon by a projection mechanism.
The latter is achieved by  coupling the new variables 
to a U(1) worldline gauge field, with an additional 
Chern-Simons  coupling set to achieve projection precisely onto the required Hilbert subspace. 
At the end, the gauge fixing of the additional U(1) gauge field leaves just an integration 
over an angle $\phi$, a modulus that describes gauge invariant configurations.
In the following, we use the variable $z=e^{\ii \phi}$, and the modular integration is obtained by 
integrating $z$ over the unit circle of the complex plane.
This construction has been exemplified in \cite{Bastianelli:2005vk,  Bastianelli:2005uy, Bastianelli:2013pta} 
and \cite{ Ahmadiniaz:2015xoa}  for worldlines with the topologies of a loop and an interval, respectively
(see also \cite{Bastianelli:2021rbt} for a pedagogical description in the case of the bi-adjoint particle).  
Eventually, the net effect of introducing the auxiliary variables $Q_a$ and $\bar Q^a$
with their U(1) coupling is to bring the dressed propagator of the photon into the form
\begin{align}
D(x_{0},y_{0},u,\bar{u};g) = \oint \frac{\dd z}{2\pi \ii} \frac {e^{z \bar{u}\cdot u} }{z^{2}} \int_{0}^{\infty} \dd T \int_{x(0)=x_{0}}^{x(1)=y_{0}}\D x \int_{\lambda(0)=0}^{\bar{\lambda}(1)=0} D \lambda D \bar{\lambda} \ e^{\ii S},
\end{align}
where the auxiliary variables have been decomposed as
\begin{align}
\bar{Q}^{a}(\tau) = z \bar{u}^{a}+ \bar{\lambda}^{a}(\tau), \qquad  Q_{a}(\tau) = u_{a}+\lambda_{a}(\tau)
\label{auxiliary-expansion-z}
\end{align}
with $\lambda_a$ and $\bar \lambda^a$ denoting the quantum fluctuations,
and with the remaining classical parts  $u_{a}$ and $\bar{u}^{a}$ describing the initial and final 
polarization of the photon depending on the modulus $z$ as indicated.
It requires a worldline action that now reads
\begin{align}\label{w-ph-action}
S = \int_{0}^{1} \dd\tau \Big( - \frac{1}{4T}g_{\mu\nu}\dot{x}^{\mu}\dot{x}^{\nu} + \ii
\bar{\lambda}^a \dot{\lambda}_a -\frac{1}{2} \dot{x}^{\mu}\omega_{\mu}^{cd}(S_{cd})_{a}\,^{b} \bar{Q}^a Q_b  
+T R_{a}\,^{b} \bar{Q}^a Q_b   -\frac{3}{4}TR
\Big).
\end{align}
The splitting of the auxliliary variables
into background and fluctuations is analogous to the splitting of the path $x^\mu(\tau)$
into a classical part plus quantum fluctuations satisfying Dirichlet boundary conditions performed earlier.
Note that in Ref.~\cite{Bastianelli:2013tsa} this precise worldline model (but with fermionic auxiliary variables and treated 
only on a loop) appears as the one corresponding to the ghost sector of gravity covariantly quantized.

\subsubsection{Examples}
The relation  between the dressed propagator and the on-shell amplitude given in Sec.~\ref{review-on-shell} is not the 
only way to obtain scattering data from the dressed propagator. Arguably the simplest way to obtain scattering data is to consider the LSZ procedure already at the level of the dressed propagator which leads to tree-level Feynman diagrams.  
After applying the LSZ procedure, the  dressed propagator in Eq.~\eqref{prop} leads to the the sum of Feynman diagrams whose topology corresponds to gravitons attached to the matter line  as depicted in  Fig.~\ref{dress-prop}. Similarly,
the simplest  cases of the dressed photon propagator are those where there is no graviton attached to the photon line and the case where a single graviton is attached to the photon line, or in other words the free photon propagator and the 3-point vertex, respectively. It is instructive to consider
these cases separately to see the role of the auxiliary variables and the counterterms in the action
\eqref{w-ph-action} before going to the semi-classical matching.
\subsubsection*{Free photon propagator}
The free photon propagator is obtained by switching off the interactions in \eqref{w-ph-action} and 
setting $g_{\mu\nu}= \eta_{\mu\nu}$.
Then,  the worldline action reduces to  
\be \label{s-free}
S_{\text{free}}=\int_{0}^{1} \dd\tau \Big( - \frac{1}{4T}\eta_{\mu\nu}
\left(\dot{x}^{\mu}\dot{x}^{\nu}
+ a^{\mu} a^{\nu} + b^{\mu}c^{\nu}
\right) + 
\ii\bar{\lambda}^a \dot{\lambda}_a
\Big).
\ee
Expanding the coordinates as $x = x_{0} + (y_{0} - x_{0})\tau + q(\tau)$,
the free path integral on quantum fluctuations produces  the measure   $(4\pi T)^{-2}$, while the classical trajectory has the 
simple action $S_{\text{cl}}= -(x_0-y_0)^2/4T$, and we are left with

\ba
D(x_{0},y_{0},u, \bar u; g)&=
\oint \frac{\dd z}{ 2 \pi \ii} \frac{e^{z u \cdot \bar u}}{z^2}\int_{0}^{\infty}\frac{\dd T } {(4\pi T)^{2}}\
e^{-\ii \frac{(x_0-y_0)^2}{4T}} = \bar u^{\mu} \Delta_{\mu\nu} (x_0-y_0) u^{\nu},
\ea
which leads to the causal photon propagator in Feynman gauge once stripping off the auxiliary variables, i.e.,
\begin{align}
\Delta_{\mu\nu} (x-y)= \eta_{\mu\nu} \int_{0}^{\infty}\frac{\dd T } {(4\pi T)^{2}}\
e^{-\ii \frac{(x-y)^2}{4T}}
=
\int \hat \dd^{4} p \frac{\ii \eta_{\mu\nu}}{p^{2} } e^{ \ii p\cdot(x-y)}.
\end{align}
\subsubsection*{Photon-photon-graviton vertex}
Let us now consider the more interesting case of a single graviton. 
In this case, we have to take into account quantum fluctuations, which we implement by performing a background
expansion of the metric tensor and the configuration space variables as 
\begin{align}
g_{\mu\nu}(x(\tau)) =\eta_{\mu\nu}+\kappa \varepsilon_{\mu\nu}(\ell) e^{\ii \ell\cdot x(\tau)},
\qquad x^{\mu}(\tau) = x_{0}^\mu + (y_{0} - x_{0})^{\mu}\tau+ q^{\mu}(\tau)
\end{align} 
with the quantum fluctuations $q^{\mu}(\tau)$ acquiring DBC, i.e. $q^{\mu}(0) = q^{\mu}(1)=0$. Notice that we have Fourier expanded the graviton field as a unique plane wave to insert just one graviton in a photon line with momentum $\ell$.
The contributions to the three-point amplitude arise from the interactions which we organize as follows:
\begin{align}
S_{\text{kin}}:=&  
\int_{0}^{1} \dd\tau \Big( - \frac{1}{4T}h_{\mu\nu}(x) 
\left(
\dot{x}^{\mu}\dot{x}^{\nu} + 
a^{\mu}a^{\nu} + b^{\mu} c^{\nu}
\right) \Big),
\\
S_{\text{spin}}:=& -\frac{1}{2} \int_{0}^{1} \dd\tau \dot{x}^{\mu}\omega_{\mu}^{cd}(S_{cd})_{a}\,^{b} \bar{Q}^a Q_b,
\label{S-spin}\\
S_{\text{Ric}}:=& T\int_{0}^{1} \dd\tau  R_{a}\,^{b} \bar{Q}^a Q_b \label{S-ric},\\
S_{\text{ct}}:=& -\frac{3}{4}T \int_{0}^{1} \dd\tau R \label{S-ct}.
\end{align}
Except for $S_{\text{kin}}$ the remaining interactions have to be background-field expanded up to $\mathcal O(\kappa)$ in order
to give contributions to the three-point amplitude.
Next, we rewrite the dressed propagator in momentum space 
introducing the external momenta of the photons $p_{1},p_{2}$, so that 
\ba
&\widetilde{D}(p_{1},p_{2},u,\bar u) = 
\int \dd^{d}x_{0} \dd^{d}y_{0} \, e^{\ii p_{1} \cdot x_{0} }e^{-\ii p_{2} \cdot y_{0}} D(x_{0},y_{0},u,\bu;g) \\
&\qquad =\hat \delta\left(p_{1} -p_{2}+\ell \right)\,  \oint \frac{\dd z}{ 2 \pi \ii} \frac{e^{z u \cdot \bar u}}{z^2}
\int_{0}^{\infty}\frac{\dd T}{(4\pi T)^{\frac{d}{2}}}e^{-\frac{\ii \xi^{2}}{4T} } \int \dd^{d}\xi \, e^{ -\ii \xi \cdot p_{2}}\, 
\,  \La e^{\ii (S_{\text{kin}}+
	S_{\text{spin}} +S_{\text{Ric}}+S_{\text{ct}} 
	) }\Ra,   
\ea
where  the expectation value evaluated on the free theory, i.e., w.r.t the action \eqref{s-free}. In the 
second equality we have performed the change of variables $2x_{+} = y_{0}+x_{0},\xi = y_{0}-x_{0}$ to factor out the momentum conservation delta function. We can now perform the perturbative expansion of the path integral\footnote{In order to perform such a task one need to use the DBC propagator $\del(\tau,\sigma)= (\tau-1)\sigma \theta(\tau-\sigma) + (\sigma-1)\tau\theta(\sigma-\tau)$ alongside with the two point function of the auxiliary variables $\la \lambda_{a}(\tau) \bar \lambda^{b} (\sigma)\ra =\delta_a\,^b \theta(\tau-\sigma)$, with $\theta(0)=1/2$.}. After  evaluating all the contributions from each term we strip-off the auxiliary variables 
as in the free photon propagator example and amputate the external legs (see also Ref.\cite{Ahmadiniaz:2015xoa} for a similar 
treatment in the case of Yang-Mills).

We list here the final results of the on-shell procedure for each contribution from the perturbative expansion.
For the kinetic term (including the regulating ghost interaction) and the spin connection vertex,
we obtain  
\be 
A_{\text{kin}}^{\alpha\beta} = -\frac{\ii \kappa}{4} \left(	p_{1} + p_{2}	\right)^{\mu} \left(	p_{1} + p_{2}	\right)^{\nu} \varepsilon_{\mu\nu}(\ell)\,\eta^{\alpha\beta}, \quad 
A_{\text{spin}}^{\alpha\beta} =\frac{\ii \kappa}{2} \left(	p_{1}^{\beta}\eta^{\alpha(\mu}p_{2}^{\nu)} + p_{2}^{\alpha}\eta^{\beta(\mu}p_{1}^{\nu)}		\right) \varepsilon_{\mu\nu}(\ell), 
\ee
respectively. The Ricci tensor vertex contribution reads
\ba 
A_{\text{Ric}}^{\alpha\beta} = \frac{\ii\kappa}{2} \Big[ -&\eta^{\mu\nu}\big(p_{1}^{\beta}p_{2}^{\alpha}+s_{12}\eta^{\alpha\beta}		\big)
+2\eta^{\alpha\beta}p_{1}^{(\mu}p_{2}^{\nu)}\\
+&p_{1}^{\beta}\eta^{\alpha(\mu}p_{2}^{\nu)}+p_{2}^{\alpha}\eta^{\beta(\mu}p_{1}^{\nu)}-s_{12} \eta^{\alpha(\mu}\eta^{\nu)\beta}
\Big]\varepsilon_{\mu\nu}(\ell),
\ea
where $s_{12} = 2p_{1}\cdot p_{2}$. Finally, the counter-term vertex generates 
\be
A_{\text{ct}}^{\alpha\beta} = \frac{3\ii\kappa}{4}\left(	s_{12}\eta^{\mu\nu}\eta^{\alpha\beta}-2\eta^{\alpha\beta}p_{1}^{(\mu}p_{2}^{\nu)}			\right)\varepsilon_{\mu\nu}(\ell).
\ee
In all of these contributions the free indices contract the external photon polarization vectors. In this
way, adding the above terms we obtain the three point vertex
\ba\label{amp}
\raisebox{-12mm}{\phg }=\frac{\ii \kappa}{2}  \, \Big(&
\frac{1}{2} s_{12}\eta^{\alpha\beta} \eta^{\mu\nu} -s_{12}\eta^{\alpha(\mu}\eta^{\nu)\beta} -2 \eta^{\alpha\beta}p_{1}^{(\mu}p_{2}^{\nu)} \\[-2em]
&+2 p_{1}^{\beta} p_{2}^{(\mu}\eta^{\nu)\alpha} +2 p_{2}^{\alpha}p_{1}^{(\mu}\eta^{\nu)\beta}  -\eta^{\mu\nu} p_{1}^{\beta} p_{2}^{\alpha}
\Big),
\ea
where the  momentum conservation delta function has been stripped off. Higher 
point examples follow a similar procedure  but with a more  involved structure. However it is obvious already that 
the dressed propagator contains much more information than required in the classical limit. 
Moreover, it raises the question of what is the role, if any,  of the auxiliary variables in the classical limit. Remember that
they were introduced only to keep track of time-ordering which is requirement forced upon us by the quantization
via path integrals. The key element to answer this question is the matching between the scattering 
amplitude and a pair of dressed propagators which we discuss next.

\subsection{From dressed propagators to amplitudes}
\label{photon-to-amplitude}
Following Ref.~\cite{Fabbrichesi:1993kz} the \emph{semi-classical} amplitude, i.e., 
the scattering amplitude with no diagrams with intermediate scalar loops, can be obtained as follows. Let us consider
the correlator 
\ba
\la  \Omega | \textrm{T} A_{\mu}(x_{1})A_{\nu}(x_{2})\varphi(x'_{1})\varphi^{\dagger}(x'_{2})|\Omega\ra 
=\mathcal{N}^{-1}\int\D g \D A \D \varphi \D \varphi^{*} e^{\ii S} A_{\mu}(x_{1})A_{\nu}(x_{2})\varphi(x'_{1})\varphi^{*}(x'_{2}), 
\ea 
where $\mathcal{N}$ is some normalization constant and $S=S_g+ S_{\text{m}}$. Remember that the 
semi-classical contributions to the scattering amplitude are obtained by disregarding diagrams with scalar and graviton loops 
(including graviton-photon loops), which is the usual  $\hbar$ power counting in path integrals. In this way, 
the integration over the scalar and the gauge fields generates the dressed propagators, namely
\be
\la  \Omega | \textrm{T} A_{\mu}(x_{1})A_{\nu}(x_{2})\varphi(x'_{1})\varphi^{\dagger}(x'_{2})|\Omega\ra =  {\cal \tilde N}^{-1}\int  \D g  \, e^{\ii S_{g}}
D_{\mu\nu}[x_{1},x_{2};g] G[x'_{1},x'_{2};g].
\ee
In order to define the $S$-matrix we need  the on-shell amputated propagator, which is constructed by the 
same prescription as in the scalar case, i.e., 
\be 
D^{c}(p,p', u, \bar u;g) := \lim\limits_{p^{2},{p'}^2 \to 0} (\ii p^{2} \ \ii p^{'2}) 
\int \dd^{4}x_{0} \dd^{4}y_{0} \, e^{\ii p\cdot x_{0} -\ii p'\cdot y_{0}} D(x_{0},y_{0}, u, \bar u;g).
\ee
The overall effect of this procedure can then again be obtained by the prescription of  replacing integration region for $\tau$ from $(0, T)$ to  $(-\infty, \infty)$  as in Ref.\cite{Mogull:2020sak}. The last step is to introduce physical 
polarization vectors, which so far have been kept track thanks to the auxiliary  variables $u$ and $\bar u$  (see previous examples). This can be achieved by setting
\begin{align}
u_\rho \bar u_\sigma \to \varepsilon_{\rho}^*{}^{} (p) \varepsilon^{}_{\sigma}(p'),
\label{polarization}
\end{align}
and therefore the  scattering amplitude in the semi-classical limit can be written as 
\ba \label{4-pt-amp}
{\cal A}\left(\gamma \varphi \to \gamma\varphi \right) = &{\cal \tilde N}^{-1}\int \left(\prod\limits_{i=1}^2 \dd x_i
\dd {x'}_i\right)
e^{-\ii p_{1}x_{1} + \ii p_{2}x_{2}} 
e^{-\ii p'_{1}x'_{1}+\ii p'_{2} x'_{2}}  \\
&\times \int \D g \, e^{\ii S_{\text{EH}}}  \varepsilon^*{}^{\mu}(p_{1}) \varepsilon_{\nu}(p_{2}) {D^c}_{\mu}\,^{\nu}[x_{1},x_{2};g] G^c[x'_{1},x'_{2};g] 
\ea
where we have rewritten the dressed photon propagator in flat space-time using  tetrad fields and at this point 
we will not distinguish between indices.

The above formula does not give yet the  contributions required to study classical light bending since we still need to take consider the geometric optics regime. 
The reason is that to obtain the classical limit of amplitudes with spin an expansion of polarization vectors in powers of $\hbar$ is still required \cite{Maybee:2019jus}. The relation \eqref{4-pt-amp} integrates out loops but does not produce an expansion of the polarization vectors which only appear after applying \eqref{polarization}. The geometric optics regime  corresponds to the situation in which we may simply replace $\varepsilon^{}_{\nu}(p_{2})$ by $\varepsilon^{}_{\nu}(p_{1})$, i.e., where the polarizations remain the same after the scattering event \cite{Cristofoli:2021vyo}.
It is well-known that in this regime, up to factors, the scattering amplitude thus obtained is equivalent to the 
amplitude between one massless and one massive scalar, see e.g. \cite{Bjerrum-Bohr:2017dxw}. 
Assuming this equivalence,  one could build up the  WQFT for  a massless scalar and a massive one by taking the massless limit of one of the dressed propagators in \eqref{amp-to-path} but this is of no interest to us. Instead we are interested in  starting from first principles building up the dressed photon propagator  and deriving such equivalence. In fact, this constitutes a consistency check of our setup.

\subsection{Partition function and derivation of Feynman rules}
Once the relation to on-shell scattering amplitudes has been made the WQFT formalism instructs us to build 
up the  following partition function
\be \label{partition-function}
\mathcal{Z}[u, \bar u]
= \oint 
\frac{\dd z}{2\pi \ii} \frac{e^{z \bar u \cdot u}  }{z^{2}}
\int \D g_{\mu\nu} \int \D x_{1} \int  \D x_{2}  \int D \lambda \int  D \bar \lambda 
\,  e^{\ii S},
\ee
where the full action reads
\begin{align}
S=S_{g} [g] + S_{\text{p}_{1}}[x_{1},\lambda,\bar \lambda,z;g] + S_{\text{p}_{2}}[x_{2};g],
\end{align}
and $S_{g}$ was defined in Eq.~\eqref{action-grav-gauge}. After rescaling the worldline parameter as 
$\tau \to \sigma/m$ we write the scalar  massive point particle action as
\begin{align}
S_{\text{p}_{2}}[x_{2};g] = - \int_{-\infty}^{+\infty} \dd\sigma \, \frac{1}{2} g_{\mu\nu}\dot{x}_{2}^{\mu}\dot{x}_{2}^{\nu}.
\end{align}
Now the point particle action can be read off from the dressed propagator of the photon in a gravitational background and 
gives
\begin{align}
\label{action}
S_{\text{p}_{1}}[x_{1},\bar{\lambda},\lambda;g] = -\int_{-\infty}^{+\infty}\dd\sigma  \left( 
\frac{1}{2}g_{\mu\nu}\dot{x}_{1}^{\mu}\dot{x}_{1}^{\nu} -\ii\bar{\lambda}\cdot \dot{\lambda}+ \frac{1}{2}\dot{x}_{1}^{\mu}\omega_{\mu}^{cd} (S_{cd})_{a}\,^{b} \bar{Q}^a Q_b 
\right) .
\end{align} 
We have also rescaled the worldline parameter of the photon action
\begin{align}
\tau = \frac{ \sigma}{ E}
\label{rescaling}
\end{align}
introducing a parameter $E$ with dimension of energy, whose physical meaning will become clear later on. For the photon worldline, we have neglected the $R_{a}\,^{b}$ vertex both with the counter-terms since they are
sub-leading in the classical limit (see example below). 
Notice here that the integration regions are now from $\sigma \in (-\infty, +\infty)$ in agreement with 
the  procedure to set dressed propagators on-shell. The partition function can now be solved perturbatively by performing wick contractions.  This task can be implemented by deriving worldline Feynman rules which take care of the perturbative expansion. 

Finally,  we need to integrate over the modulus $z$ and consider the geometric optics regime. This
regime can be applied in a similar way we did in the amplitude case using Eq.~\eqref{polarization} 
but we find more 
convenient to consider it directly from the partition function by defining
\begin{align}
\mathcal Z_{\text{geom-opt}}:=-
\mathcal{Z}(u, \bar u)\Big|_{ \bar u  \to u },
\end{align}
where the minus sign is due $u \cdot  \bar u=-1$, which is consistent with our metric signature.  This definition incorporates the geometric optics limit by disregarding terms of $\mathcal{O}(\hbar)$ coming from polarization vectors as discussed at the end of last Section. In addition, without loss of generality, we have chose a basis of real polarization vectors
since in that limit any choice leads to the same result.   

The Feynman rules to keep track of the perturbative expansion and obtain the partition function are now
easy to derive. First,  performing a background  expansion of  the tetrads and the spin connection alongside the metric tensor  we have 
\begin{align}
e^{a}_{\mu} &= \eta^{a}_{\mu}+\frac{\kappa}{2}h^{a}_{\mu} - \frac{\kappa^{2}}{8}h_{\mu\alpha}h^{\alpha a} + \mathcal{O}(\kappa^{3}),\\
\omega_{\mu}\,^{ab} &= -\kappa \partial^{[a}h^{b]}_{\mu}-\frac{\kappa^{2}}{2}h^{\nu[a}\left(\partial^{b]}h_{\mu\nu} -\partial_{\nu}h^{b]}_{\mu} + \frac{1}{2}\partial_{\mu}h^{b]}_{\nu}
\right) + \mathcal{O}(\kappa^{3}),  
\end{align}
which can be inserted in the action in \eqref{action} alongside with the background expansion of the configuration space variables 
\be 
x^{\mu}(\sigma)=b^{\mu}+p^{\mu}\sigma + q^{\mu}(\sigma)\;
\ee
and auxiliary variables as in Eq.~\eqref{auxiliary-expansion-z}.
Notice that the momentum $p^{\mu}$ is massless for the case of the photon.  The Fourier transform of  the fluctuations of the auxiliary variable are given by
\begin{equation}
\lambda_{a}(\sigma)=\int \hat \dd \omega \, e^{-\ii \omega \sigma} \lambda_{a}(\omega),
\quad  \bar \lambda^{a}(\sigma)=\int \hat \dd \omega \, e^{\ii \omega \sigma} \bar \lambda^{a}(\omega).
\end{equation}
Finally,  Fourier expanding the graviton field as
\begin{align}
h_{\mu\nu}(b+p\tau+q) = \sum_{n=0}^{\infty}\frac{(-\ii)^{n}}{n!} \int \hat \dd^4 \ell \, e^{-\ii\ell\cdot (b+p\tau)} \left( q(\tau)\cdot \ell\right)^{n}\, \varepsilon_{\mu\nu}(\ell)  
\end{align}
and plugging the above expansion in the action, we can obtain the Feynman rules associated with the photon
worldline. From now on we will not distinguish between curved and tangent indices. 

The 2-point 
functions associated with the fluctuations lead to the photon worldline propagator  
\begin{align}
\raisebox{-1.7mm}{
	\begin{tikzpicture}[thick]
	\coordinate (A) at (-1,-0);
	\coordinate (B) at (1,-0);
	\filldraw (A) circle (2pt) node[left] {$q^\mu$};
	\filldraw (B) circle (2pt) node[right] {$q^\nu$};
	\draw[->,>=stealth] (-0.5,0.2) -- (0.5,0.2) node[midway,above] {$\omega$};
	\path[very thick, draw,snake it] (-1,0)--(1,0); 
	\end{tikzpicture}
}
= - \ii\frac{ \eta^{\mu \nu}}{2}\left( \frac{1}{(\omega+\ii\epsilon)^2}+\frac{1}{(\omega-\ii\epsilon)^2} \right),
\end{align}
where, as in the case of the massive worldline, we use time-symmetric propagators. The Feynman rules associated with the photon worldline interactions read
\begin{align}\label{Frule-LO-photon}
\raisebox{-10mm}{\begin{tikzpicture}[thick]
	\path [dashed, draw,snake it]
	(-1,-1) -- (1,-1);
	\path [double, draw,snake it]
	(0,-1) -- (0,-2)node[below]{$h_{\mu\nu}$};
	\draw[->,>=stealth] (-0.5,-1.2) -- (-0.5,-1.8) node[midway, left]
	{$k$};
	\coordinate (A) at (0,-1);
	\filldraw (A) circle (1.5pt);
	\end{tikzpicture}}&=\frac{\ii \kappa}{2}  e^{\ii k \cdot b}\hat\delta(k\cdot p)\left(
-p^{\mu}p^{\nu} + \ii z k^{\alpha} p^{(\nu}\eta^{\mu)\beta}(S_{\alpha\beta})_{\rho}\,^{\sigma} \bar u^{\rho} u_{\sigma}
\right) ,\\[-2em] \nonumber \\ 
\raisebox{-10mm}{
	\begin{tikzpicture}[thick]
	\path [dashed, draw,snake it]
	(-1,-1) -- (0,-1);
	\path [very thick,draw,snake it]
	(0,-1) -- (1,-1) node[right]
	{$q^\rho(\omega)$};
	\path [double, draw,snake it]
	(0,-1) -- (0,-2)node[below]{$h_{\mu\nu}$};
	\draw[->,>=stealth] (-0.5,-1.2) -- (-0.5,-1.8) node[midway, left]
	{$k$};
	\coordinate (A) at (0,-1);
	\filldraw (A) circle (1.5pt);
	\end{tikzpicture}}&= 
\frac{\kappa}{2}e^{\ii k\cdot b} \hat \delta\left(k\cdot p + \omega\right) \left[ \left(
p^{\mu}p^{\nu}k_{\rho}+2\omega p^{(\mu}\delta^{\nu)}_{\rho}
\right)  \right.\\[-1em]
&\hskip 1.5cm \left. -\ii z k^{\alpha} \left(	\eta^{\beta (\mu} (p^{\nu)} k_{\rho} +\omega \delta^{\nu)}_{\rho})	\right)
(S_{\alpha\beta})_{\lambda}\,^{\delta} \bar u^{\lambda} u_{\delta}
\right] 
,\nonumber
\end{align}
where the solid photon line represents a fluctuation of the worldline photon line. At this order we also have 
vertices related to the fluctuations of auxiliary variables which we obtain by expanding
the spin connection vertex  in \eqref{action}  as well, leading to
\begin{align}
\raisebox{-10mm}{ \begin{tikzpicture}[thick]
	\path [dashed, draw,snake it]
	(-1,-1) -- (0,-1);
	\path [dashed, draw]
	(0,-1) -- (1,-1);
	\path [double, draw,snake it]
	(0,-1) -- (0,-2)node[below]{$h_{\mu\nu}$};
	\draw[->,>=stealth] (-0.5,-1.2) -- (-0.5,-1.8) node[midway, left]
	{$k$};
	\path [draw]
	(0,-1) -- (1,-1) node[right]
	{$\lambda_{\sigma}(\omega)$};
	\draw[->,>=stealth] (0.5,-1) -- (0.6,-1);
	\coordinate (A) at (0,-1);
	\filldraw (A) circle (1.5pt);
	\end{tikzpicture}}&= -\frac{\kappa}{2} e^{\ii k \cdot b} \hat \delta(\omega+ k \cdot p)
z \,	k^{\alpha} p^{(\nu} \eta^{\mu)\beta}(S_{\alpha \beta})_{\rho}\,^{\sigma}\bar u^{\rho}	,
\\
\raisebox{-10mm}{\begin{tikzpicture}[thick]
	\path [dashed, draw,snake it]
	(-1,-1) -- (0,-1);
	\path [dashed, draw]
	(0,-1) -- (1,-1);
	\path [double, draw,snake it]
	(0,-1) -- (0,-2)node[below]{$h_{\mu\nu}$};
	\draw[->, >=stealth] (-0.5,-1.2) -- (-0.5,-1.8) node[midway, left]
	{$k$};
	\path [draw]
	(0,-1) -- (1,-1) node[right]
	{$\bar{\lambda}^{\rho}(\omega)$};
	\draw[<-,>=stealth] (0.5,-1) -- (0.6,-1);
	\coordinate (A) at (0,-1);
	\filldraw (A) circle (1.5pt);
	\end{tikzpicture}}&= 
-\frac{\kappa }{2} 
e^{\ii k\cdot b}\hat \delta( k\cdot p-\omega) k^{\alpha} p^{(\nu} \eta^{\mu)\beta} (S_{\alpha\beta})_{\rho}\,^{\sigma}u_{\sigma},
\end{align}
where the arrow distinguishes between $\bar{\lambda}^{a}$ and $\lambda^{a}$ which scale differently with $z$. Other 
rules can be easily derived, e.g.,  the  rule for a 2-fluctuation reads
\ba
\raisebox{-15mm}{
	\begin{tikzpicture}[thick]
	\path [dashed, draw,snake it]
	(-1,-1) -- (0,-1);
	\path [very thick,draw,snake it]
	(0,-1) -- (1,-1)node[right]
	{$q^\rho(\omega_{1})$};;
	\path [very thick,draw,snake it]
	(0,-1) -- (1,0)node[right]
	{$q^\sigma(\omega_{2})$};;
	\path [double, draw,snake it]
	(0,-1) -- (0,-2)node[below]{$h_{\mu\nu}$};
	\draw[->,>=stealth] (-0.5,-1.2) -- (-0.5,-1.8) node[midway, left]
	{$k$};
	\coordinate (A) at (0,-1);
	\filldraw (A) circle (1.5pt);
	\end{tikzpicture}
}=&
\frac{\ii \kappa}{2}e^{\ii k\cdot b}\hat \delta \left(k\cdot p+\omega_{1} + \omega_{2} \right) \left[
\frac{1}{2}p^{\mu}p^{\nu} k_{\rho}k_{\sigma} + \omega_{1}k_{\sigma} p^{(\mu}\delta^{\nu)}_{\rho}
+ \omega_{2}k_{\rho} p^{(\mu}\delta^{\nu)}_{\sigma} 
\right.
\\[-2em]
&\left.+ \omega_{1}\omega_{2} \delta^{(\mu}_{\rho}\delta^{\nu)}_{\sigma} -\frac{\ii }{2} z k^{\alpha} (S_{\alpha \beta})_{\lambda}\,^{\delta} \bar u^{\lambda}u_{\delta} \left(
p^{(\mu} \eta^{\nu)\beta} k_{\rho}k_{\sigma} + \omega_{1}\eta^{\beta (\nu}\delta^{\mu)}_{\rho} k_{\sigma}
\right)
\right].
\ea
At the order we are interested we have computed all the required Feynman rules associated with the photon worldline. In addition to the rules associated with the scalar worldline in Section~\ref{review} this is all we need to compute the eikonal phase.

\subsection{Definition of the eikonal phase and deflection angle}
\label{def-eikonal-phase}
Let us move on with the definition of the eikonal phase in the geometric optics regime. The eikonal phase is  related to the 
partition function as
\begin{align}
\mathcal Z_{\text{geom-opt}}=e^{\ii \chi},
\end{align}
and hence by the usual definition of the eikonal phase
\begin{align}
e^{\ii \chi}:= \frac{1}{4 m E} \int \hat \dd^4 q\, \hat\delta ( q \cdot v_1) \hat \delta(  q \cdot v_2) e^{\ii q\cdot b }
\mathcal{A} (\phi \gamma \to \phi \gamma),
\end{align}
we can relate the scattering amplitude  to the free energy of the WQFT in the geometric optics regime. 
Here $E$ is the energy of the photon. The rescaling
we introduced in Eq.\eqref{rescaling} then makes the job of obtaining the eikonal phase directly in terms of momenta.  We write 
\begin{align}
\ii \chi= \ii(\chi_1+ \chi_2 + \dots ),
\end{align}
where $\chi_i$ is of order $\mathcal{O}(\kappa^{2i})$. Following \cite{Bjerrum-Bohr:2016hpa} we define the deflection
angle in the small angle approximation at each order in perturbation theory by 
\begin{equation}
\theta_i=-\frac{1}{E} \frac{\partial \chi_i }{\partial |b|}.
\end{equation}
\subsubsection{Example}
\label{example-3-points}
In order to illustrate our methods let us give a simple example. 
The three-point amplitude $\mathcal{M}(\gamma \gamma h)$ is formally vanishing in the optical regime so let us consider its off-shell version $\mathcal{M}^{\mu\nu}(\gamma \gamma h)$ instead. Consider the 3-point vertex of 
\eqref{amp} and suppose we are interested in the limit where the graviton is soft.    Let  $p_{2}^{\mu} = p_{1}^{\mu}+\hbar  \bar{q}^{\mu}$ where $\bar{q}$ is interpreted in the classical limit as a wave-number related to the graviton. 
The  polarization vector thus satisfies
$\varepsilon_{\mu}(p_{2}) = \varepsilon_{\mu}(p_{1})+ \mathcal{O}(\hbar)$. We also parameterize  the incoming photon momentum as $p_{1}= Ev^{\mu} =(E,0,0,E)$ where $E$ is the energy of the photon. Then, up to terms proportional of 
$\hbar$, we obtain
\be 
{\cal M}^{\mu\nu}(p_{1},p_{2}) = \ii \kappa  \,  p_{1}^{\mu}p_{1}^{\nu} +\mathcal{O}(\hbar) =
\ii\kappa E^{2} v^{\mu} v^{\nu} +\mathcal{O}(\hbar),
\label{amp-3-point}
\ee
where we have used the on-shell condition on the momenta $p_{1}$ and $p_{2}$ and the
transversality of the photon polarization vectors. 

The equivalent object in WQFT is obtained from the LO Feynman rule \eqref{Frule-LO-photon}. The partition
function then simply reads
\begin{align}
{\cal Z}(z, u ,\bar u)= -\frac{\ii \kappa}{2}\int \, \hat \dd^4 q \, e^{\ii q \cdot b}\deltahat(q\cdot p)
\left[p^{\mu }p^{\nu }  + 2 z p^{\nu}(q\cdot \bar{u} \,u^{\mu }-\bar u^{\mu}\,  q\cdot u) \right].
\end{align}
Clearly, in order to match this expression with the previous it must be that the extra terms must vanish. To see that this 
is the case let us consider the integration of the kernel over $z$. Exchanging the integration orders and using the identity 
\begin{equation}
\begin{aligned}
\frac{1}{2\pi \ii} \oint  \dd z \frac{e^{z \bar u \cdot u}  }{z^{k+1}}=  
\begin{cases}
\frac{ (u \cdot \bar u)^k}{k!}, & k \ge 0,\\ 
0, & \text{otherwise} ,
\end{cases}
\label{identity-auxiliary-variables} 
\end{aligned}
\end{equation}
we obtain 
\begin{align}
{\cal Z}(z, u ,\bar u)= - \frac{\ii \kappa}{2}u_\rho \bar u_\sigma
\int  \hat \dd^4 q \, e^{\ii q\cdot b}\deltahat(q\cdot p)
p^{\mu }  \left[p^{\nu } \eta^{\rho\sigma} -2\ii q_{\lambda} (S^{\lambda \nu})^{\rho \sigma} 
\right].
\end{align}
Notice that the dependence on $(\bu,u)$ on the first term appears due to this identity. 
Furthermore, using the anti-symmetry of $(S^{\lambda \nu})_{\rho}^{\ \sigma}$ yields to the following relation
\be \label{lop}
{\cal Z}_{\text{geom-opt}}^{\mu\nu} = \int \hat \dd^{4}q\, e^{\ii q\cdot b}\,\deltahat(q\cdot p)\left(-\frac{\ii \kappa}{2}p^{\mu}p^{\nu}	\right) ,
\ee
where we stripped off the graviton polarization vector. Therefore, up to an irrelevant sign,  we see that in the geometric optics regime the leading order Feynman rule matches the classical limit computed in Eq.~\eqref{amp-3-point} 
as expected.

\subsection{Massless limit vs photon worldline in WQFT}
\label{equivalence-massless}
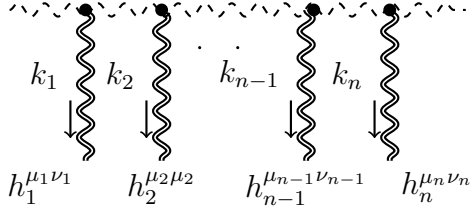
\begin{figure}[htb]
	\centering
	\begin{tikzpicture}[thick]
	\draw[double, draw, snake it] (-2,0)--(-2,-2)node[below left]
	{$h_1^{\mu_1 \nu_1}$};
	\draw[<-,,<-=stealth, draw] (-2.2,-1.7)--(-2.2,-1.2) node[ above left]
	{$k_1$};
	\draw[double, draw, snake it] (-1,0)--(-1,-2)node[below]
	{$h_2^{\mu_2\mu_2}$};
	\draw[<-,,<-=stealth, draw] (-1.2,-1.7)--(-1.2,-1.2) node[ above left]
	{$k_2$};
	\draw[double, draw, snake it] (0.9,0)--(0.9,-2)node[below]
	{$h_{n-1}^{\mu_{n-1}\nu_{n-1}}$};
	\draw[<-,,<-=stealth, draw] (0.7,-1.7)--(0.7,-1.2) node[above left]
	{$k_{n-1}$};
	\draw[double, draw, snake it] (2,0)--(2,-2)node[below right]
	{$h_n^{\mu_n \nu_n}$};
	\draw[<-,,<-=stealth, draw] (1.8,-1.7)--(1.8,-1.2) node[above left]
	{$k_{n}$};
	\path [dashed, draw, snake it]
	(-3,0) --(0,0) -- (3,0);
	\path [dashed, very thick, dots=2 per 1cm, draw](-0.5,-0.5) -- (0.5,-0.5);
	\coordinate (A) at (-1,0);   \filldraw (A) circle (2.pt);
	\coordinate (B) at (-2,0);   \filldraw (B) circle (2.pt);
	\coordinate (C) at (1,0);   \filldraw (C) circle (2.pt);
	\coordinate (D) at (2,0);   \filldraw (D) circle (2.pt);
	\end{tikzpicture}
	\caption{Dressed propagator with external photon lines on-shell}
	\label{dress-prop-photon}
\end{figure}

On general grounds we expect that  the contributions related to the spin-terms should not contribute in the geometric optics 
regime since assuming the equivalence principle one may replace the photon worldline by the worldline of a massless scalar.
We had a taste of this in the calculation of the lowest order partition function
\eqref{lop}. For the general case let us consider the diagram shown in Fig.~\ref{dress-prop-photon} which appears as a sub-topology of our worldline diagrams. Let $\mathcal{K}$  denote the mathematical expression of this diagram and let  $k_i$ be the momenta associated with each graviton line which we will consider as outgoing. The momentum of the photon worldline is
labeled  by $p^\mu$.
This diagram is still to be integrated over $z$ and hence the identity \eqref{identity-auxiliary-variables} implies that after integration all terms proportional to $z^i$ vanish for $i>1$. The same identity  produces the factor $u\cdot \bar u$, which gives a non-vanishing contribution associated with the term independent of $z$. In other words only the terms independent of $z$ and proportional to $z$ contribute. Now let us rewrite the Feynman rule of a single vertex in the form
\begin{align}
V_i^{\mu_1\nu_1}(p,k_i):= \frac{\ii \kappa }{2} e^{-\ii k_i \cdot p} (-p^{\mu_i} p^{\nu_i}+z B_i^{\mu_i \nu_i}),
\end{align}
where 
$$B_i^{\mu_i \nu_i}:=\ii  k_{i}^{\alpha}p^{(\nu_{i}}\eta^{\mu_{i} ) \beta} (S_{\alpha\beta})_{\rho}\,^{\sigma} \bar u^{\rho} u_{\sigma}.$$
From the Feynman rules and on the support of the Dirac delta functions this diagram can be organized as follows
\begin{align}
\mathcal{K}^{\mu_1 \cdots \mu_n \nu_1 \dots \nu_n}= \frac{\ii^n}{2^{n}} \kappa^n \left(\prod_{i=1}^n \hat \delta(k_i \cdot p) e^{\ii k_i\cdot b}\right) 
K^{\mu_1 \cdots \mu_n \nu_1 \cdots \nu_n },
\label{main-formula-ngravitons}
\end{align}
where the only non-vanishing contributions are proportional to $z$ due to \eqref{identity-auxiliary-variables}. Hence 
\begin{align}
K^{\mu_1 \nu_1 \cdots \mu_n \nu_n }=(-1)^n 
p^{\mu_1} p^{\nu_1} \cdots p^{\mu_n} p^{\nu_n} -z \sum_{i=1}^{n}
p^{\mu_1} p^{\nu_1} \cdots \widehat{ p^{\mu_i} p^{\nu_i}} \cdots p^{\mu_1} p^{\nu_1}
B_i^{\mu_i \nu_i} \label{tensor-z},
\end{align}
where the hat means that the factor should be excluded. Since the coefficient $B_i^{\mu_i \nu_i}$ depends on a single 
factor of the spin tensor we can use the anti-symmetry of $(S^{\lambda \nu})_{\rho}^{\ \sigma}$ and that $\bar u \to u$ 
thus concluding that  all terms proportional to $z$ vanish the geometric optics regime as expected\footnote{For complex polarization vectors a similar proof can be devised.}.  
The same exercise
can be  done in the case where fluctuations on the photon worldlines or auxiliary variables are considered obtaining the same
result. A simple realization of this will be shown explicitly in the next Section. Let us conclude by stressing 
that it is only the combination of the geometric optics regime and the identity \eqref{identity-auxiliary-variables} that makes these terms vanishing. This provides an alternative path to show the equivalence  between the massless limit of 
scattering amplitudes involving two massive particles and the amplitudes of photons and a single massive particle in 
the classical limit, which in WQFT can be understood as disregarding the spin tensor.

Classical deviations from the geometric-optics regime --- known as the gravitational spin Hall effect\footnote{See Ref.~\cite{Oancea:2019pgm} for a review and references therein.} --- have been studied from first
principles in  Ref.~\cite{Oancea:2020khc} for the case of propagation of light and Ref.~\cite{Andersson:2020gsj} for 
propagation of gravitational waves.

\section{Calculation of deflection angles}
\label{calculations}

The WQFT setup is now complete so we are ready to apply it to perturbative calculations of the deflection angle based on 
the eikonal phase. An example of the  calculation of the momentum impulse is shown in Appendix \ref{app:B} for the 
interested reader. In Section \ref{dressed-propagator}, we have considered only the case of a spinless massive particle but it is easy to generalize it to the  case of spin. The treatment of classical spinning massive particles in WQFT has been discussed at length in Ref.~\cite{Jakobsen:2021zvh} so we will build on these  results. For the spinning case we will conform ourselves with a LO calculation up to quadratic order in spin. 

We will use the result of Sec.~\ref{equivalence-massless} which implies that the spin tensor plays no role 
in our computations. Therefore the integration over the modulus $z$ is trivial (see example in Sec.~\eqref{example-3-points}) and produces the contraction of the auxiliary variables  $\bar u \cdot u=-1$ in the geometric optics regime.
For calculations we will parametrize the momenta of the particles  as $p_1=E v_1$ and $p_2= m v_2$, where $v_1^2=0$ and $v_2^2=1$. It will also be useful to choose the rest frame of the massive particle such that $v_1=(1, 0, 0, 1)$ and $v_2=(1,0,0,0)$. Finally, recalling
that $b^\mu$ is space-like we define 
$|b| \equiv \sqrt{-b^2}$. 

\subsection{The spinless case}
\label{eikonal-spinless}
\begin{figure}
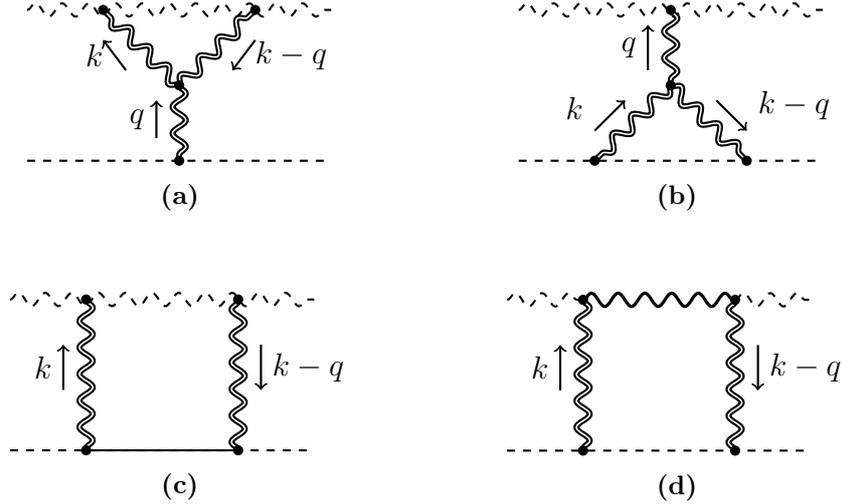

	\centering
	\begin{subfigure}{0.4\textwidth}
		\centering
		\nloeiktwo   
		\caption{}
		\label{fig:3grav1}
	\end{subfigure}
	\begin{subfigure}{0.4\textwidth}
		\centering
		\nloeikone 
		\caption{}
		\label{fig:3grav2}
	\end{subfigure}
	\vskip 1cm 
	\begin{subfigure}{0.4\textwidth}
		\centering
		\nloeikthree 
		\caption{}
		\label{fig:box1}
	\end{subfigure}
	\begin{subfigure}{0.4\textwidth}
		\centering
		\nloeikfour
		\caption{}
		\label{fig:box2}
	\end{subfigure}
	\caption{NLO diagrams}
	\label{NLO:diags}
\end{figure}

Let us now move to the evaluation of the eikonal phase and the deflection angle in the spinless case. At LO there is 
a single diagram contributing to the eikonal phase while at NLO there are four diagrams that have the correct power 
counting. The latter are shown in Fig.~\ref{NLO:diags}.
\subsubsection{Leading Order}
At this order the Feynman rules give a single diagram, and from 
our definition of the eikonal phase in Section \ref{def-eikonal-phase}, the LO eikonal phase reads
\begin{equation}
\ii \chi_1 =
\raisebox{-10mm}{
	\loeik
}
=-\ii \kappa^2  \frac{(p_1 \cdot p_2)^2 }{4 } 
\int \hat  \dd^{4}q \, \hat \delta(q \cdot p_{1})\hat \delta(q \cdot p_{2}) \frac{ e^{ \ii q\cdot b}}{q^2}. 
\;
\end{equation}
To regulate the divergent integral in this expression we find convenient to  use a cut-off regulator. The regulated 
integral then reads
\begin{align}
\mathcal  I_{\simtree}=  \int \hat \dd^4 q \frac{\hat \delta (q \cdot p_1) \hat \delta (q \cdot p_2 )}{q^2} e^{\ii q\cdot b}
=
\frac{1}{4\pi p_1\cdot  p_2} \log \left(\frac{|b|^2}{L^2} \right).
\label{I-case-simple}
\end{align}
To obtain this result we have set up a differential equation for $\mathcal  I_{\simtree}$ such that the derivative produces a 
finite expression following  similar steps as in  Ref.~\cite{delaCruz:2020bbn} with minor changes related to the 
parametrization of momenta. Remember that in $d=4$ we set  $p_1=E (1,0,0,1)$ and $p_2= m (1,0,0,0)$. Therefore we obtain
\begin{align}
\chi_1= -2 G_N (p_1 \cdot p_2) \log \left(\frac{|b|^2}{L^2} \right),
\end{align}
where we have used $\kappa^2=32 \pi G_N$. Finally, using the rest frame of the  massive particle
we have
\begin{align}
\theta_1 = -\frac{1}{E} \frac{ \partial \chi_1}{ \partial |b|}
= \frac{4 G_N m}{|b|},
\end{align}
which matches the result from general relativity.

\subsubsection{NLO}
Moving on with the NLO calculation,  let us start with the diagrams involving the 3-graviton vertex\footnote{We use the conventions of  Ref.~\cite{Bjerrum-Bohr:2014lea}. See also Ref.~\cite{Plefka:2018dpa}.}. 
\subsubsection*{Diagram (a)}
This diagram vanishes identically. In order to see this first notice that this diagram contains as a subtopology the diagram in Fig.~\ref{dress-prop-photon}.
Therefore, from Eq.~\eqref{main-formula-ngravitons} the integrand of this subtopology is proportional to 
\begin{align}
\hat \delta (p_1 \cdot q) \hat \delta (p_1 \cdot k) p_1^{\mu_1} p_1^{\mu_2} p_1^{\nu_1} p_1^{\nu_2}.
\end{align}
Upon using the 3-graviton vertex Feynman rule and $p_1^2=0$, we obtain identically zero due to the Dirac-delta
constraints. This result is independent of the other subtopology containing the matter worldline.

\subsubsection*{Diagram (b)}
This diagrams evaluates to
\begin{equation}
\raisebox{-10mm}{
	\nloeikone
}
=-\ii   \frac{\kappa^4 m^2  }{32} 
\int \hat \dd^{4}q \, \hat \delta(q \cdot p_{1})\hat \delta(q \cdot p_{2})
\frac{ e^{\ii q\cdot b}}{q^2} \int \hat \dd^4 k \, \hat \delta(k \cdot p_2)\frac{N_1(q,k,p_1)}{k^2(k-q)^2}, 
\;
\end{equation}
where $N_1(q,k,p_1)=(p_1 \cdot p_2)^2(k^2+(k-q)^2)+m^{2}(k \cdot p_1)^2$. The terms multiplying $(p_1 \cdot p_2)^{2}$ lead to tadpole integrals which are vanishing. Using our parametrization of momenta  let us focus on the integral
\begin{align}
I_1= \int \hat \dd^4 k  \hat \delta(k \cdot v_2) \frac{(k \cdot v_1)^2}{k^2(k-q)^2}= v_1^\mu v_1^\nu
\int \hat \dd^4 k  \, \hat \delta(k \cdot v_2) \frac{k_\mu k_\nu}{k^2(k-q)^2}.
\end{align}
On the support of the Dirac-delta constraints  $\hat \delta(q \cdot v_{1})\hat \delta(q \cdot v_{2})$ this integral
can be reduced by performing a simple Passarino-Veltman reduction leading to 
\begin{align}
I_1= \frac{v_1^\mu v_1^\nu}{8}(3 q_\mu q_\nu+q^2({v_2}_\mu {v_2}_\nu-\eta_{\mu\nu}))
\int \hat \dd^4 k   \frac{\hat \delta(k \cdot v_2)}{k^2(k-q)^2}=\frac{q^2 \sigma^{2}}{8}\int \hat \dd^4 k \frac{\hat \delta(k \cdot v_2)}{k^2(k-q)^2},
\end{align}
where we have defined $\sigma= v_{1}\cdot v_{2}$ to write the result in a Lorentz invariant form. Then we are left to 
calculate the integral
\begin{align}
I_\vartriangleright :=\int \hat \dd^{4}q \, \hat \delta(q \cdot v_{1})\hat \delta(q \cdot v_{2}) e^{\ii q \cdot b}
\int \hat \dd^4 k \frac{ \hat \delta(k \cdot v_2)}{  k^2(k-q)^2},
\label{integral-triangle-massless}
\end{align}
which can be computed following Ref.\cite{Bjerrum-Bohr:2018xdl} adapted to our case. The result is
\begin{align}
I_\vartriangleright= \frac{1}{16\pi |b|} .
\end{align}
Let us now move on to diagrams with the topology of a box.

\subsubsection*{Diagram (c)}
This diagram  evaluates to 
\begin{equation}
\raisebox{-10mm}{
	\nloeikthree
}
= \ii   \frac{(p_1 \cdot p_2)^{4} \kappa^{4} }{32} 
\int\hat \dd^{4}q \, \hat \delta(q \cdot p_{1})\hat \delta(q \cdot p_{2})e^{\ii b\cdot q}\int \hat \dd^4 k  \frac{\hat  \delta(k \cdot p_1)k\cdot(k-q)}{(k\cdot p_2)^2 k^2(k-q)^2}\;.
\end{equation}   
The integral reduction produces integrals with double poles on the same side of the complex plane. Following
Ref.~\cite{Kalin:2020mvi} these integrals vanish after closing the contour in the opposite direction\footnote{For time symmetric
	propagators one simple applies this argument twice for each $\ii \epsilon$ prescription.}. 

\begin{figure}
	\centering
	\begin{tikzpicture}[thick]
	\draw[double, draw, snake it] (-1,0)--(-1,-2)node[below]
	{$h_1^{\mu_1\mu_1}$};
	\draw[<-,,<-=stealth, draw] (-1.2,-1.7)--(-1.2,-1.2) node[ above left]
	{$q_1$};
	\draw[double, draw, snake it] (0.9,0)--(0.9,-2)node[below]
	{$h_{2}^{\mu_{2}\nu_{2}}$};
	\draw[<-,,<-=stealth, draw] (0.7,-1.7)--(0.7,-1.2) node[ above left]
	{$q_{2}$};
	\path [dashed, draw, snake it] (-2,0) --(-1,0);
	\path [dashed, draw, snake it](1,0) -- (2,0);
	\path [draw, snake it]    (-1,0) --(0,0) -- (1,0);
	\coordinate (A) at (-1,0);   \filldraw (A) circle (2.pt);
	\coordinate (C) at (1,0);   \filldraw (C) circle (2.pt);
	\end{tikzpicture}
	\caption{Subtopology of diagram with photon  fluctuation}
	\label{subtopo-fluctuation}
\end{figure}
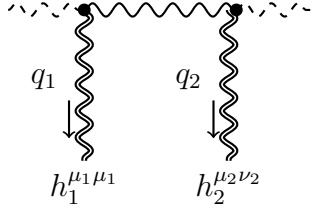

\subsubsection*{Diagram (d)}
Let us show that in this case the spin tensor does not contribute in the 
geometric optics regime. For that consider a simpler version of the exercise in Section \ref{equivalence-massless} but now including  a single fluctuation as shown in Fig.~\ref{subtopo-fluctuation}. Focusing on the terms proportional to 
$z$ we find that the  integrand is proportional to the tensor structures 
\begin{align}
\ii \eta^{\mu_1\nu_2} p_1^{\mu_2} (S_{\alpha}^{\,\, \nu_1})^{\rho \sigma} u_\rho \bar{u}_\sigma q_i^{ \alpha}, 
\qquad \ii p_1{}^{\mu_1} p_1{}^{\nu_1} q_i{}^{\nu_2} 
(S_{\alpha}^{\,\, \mu_2})^{\rho \sigma} u_\rho \bar{u}_\sigma q_i^{ \alpha}, \quad i=1,2,
\end{align}
which vanish in the geometric optics regime. The other contributions simplify to
\begin{align}
\raisebox{-10mm}{
	\nloeikfour} =\ii\frac{\kappa^4 (p_1 \cdot p_2)^{2} }{32} 
\int \hat \dd^{4}q \hat \delta(q \cdot p_{1})\hat \delta(q \cdot p_{2}) e^{\ii q\cdot b}
\int \hat \dd^4 k \frac{\hat \delta(k \cdot p_2) \, N_2(q,k,p_1)}{ (k\cdot p_1)^2 k^2(k-q)^2}, 
\;
\end{align}
where $N_2=(p_1 \cdot p_2)^2(k^2-k\cdot q)+2 m^{2}(k\cdot p_1)^2$. The integral reduction produces finite integrals with  
double poles on the same side of the complex plane which we can set to zero. Therefore the only surviving term is 
the one that cancels the double, which is proportional to \eqref{integral-triangle-massless}. 

Therefore  after adding up 
the contributing  diagrams (b) and (d) the result of the eikonal phase reads
\begin{align}
\chi_2 = \kappa^4  \frac{15}{256}  m (p_1 \cdot p_2) \frac{1}{16 \pi |b|}, 
\end{align}
and  using our parametrization of momenta the scattering angle reads
\begin{align}
\theta_2 = -\frac{1}{E}  \frac{\partial \chi_2}{ \partial |b|}
= \frac{15\pi}{4}\frac{ G_{N}^2 m^2}{|b|^2},
\end{align}
in agreement with the massless limit of the scattering angle of two massive objects in which one of the masses 
goes to zero.

\subsection{Spinning massive particle}
\label{spin}
The case of a spinning massive particle can be treated along the same lines.  The description in Ref.~\cite{Jakobsen:2021zvh}  is based on the inclusion of supersymmetry at the level of the  the worldline action. The WQFT thus 
constructed is valid at quadratic order in spin. Here we conform  ourselves with summarizing the LO Feynman rule for
this case. Performing the same rescaling  as in the previous Section we have the Feynman rule 
\begin{align} 
\label{eq:vertex-spin}
\raisebox{-10mm}{\begin{tikzpicture}[thick]
	\path [dotted, draw]
	(-1,-1) -- (1,-1);
	\path [double, draw,snake it]
	(0,-1) -- (0,-2)node[below]{$h_{\mu\nu}$};
	\draw[->,>=stealth] (-0.5,-1.2) -- (-0.5,-1.8) node[midway, left]
	{$k$};
	\coordinate (A) at (0,-1);
	\filldraw (A) circle (1.5pt);
	\end{tikzpicture}}&
= -\ii \frac{ \kappa}{2} e^{\ii k \cdot b} \hat \delta(k \cdot p) 
\left(p^{\mu} p^{\nu}+\ii m (k\cdot \mathcal{S})^{({\mu}} p^{\nu)}-
\frac{1}{2}m^{2}(k\cdot \mathcal{S})^{\mu}(k\cdot \mathcal{S})^\nu\right),
\end{align}
where $(k\cdot \mathcal{S})^\mu:=k_\nu \mathcal{S}^{\nu\mu}$. Therefore at LO the eikonal phase is computed from a 
single diagram obtaining
\begin{equation}
\ii \chi_1 =
\raisebox{-10mm}{
	\loeik}
=-\ii \kappa^2  \frac{(p_1 \cdot p_2)^2 }{4 } 
\int \hat \dd^{4}q \, \hat \delta(q \cdot p_{1})\hat \delta(q \cdot p_{2}) \frac{ e^{\ii q\cdot b} (1+N_{\mathcal{S}})}{q^2}, 
\;
\end{equation}
with the numerator $ N_{\mathcal{S}}$ is given by
\be
N_{\mathcal{S}}= -\frac{\ii m}{p_{1}\cdot p_{2}}\left(p_{1}\cdot {\cal S}\cdot q\right) -\frac{m^{2}}{2(p_{1}\cdot p_{2})^{2}}\left( p_{1}\cdot {\cal S}\cdot q
\right)^{2}
\;.
\ee
Specializing the spin tensor defined in  Ref.~\cite{Jakobsen:2021zvh} for the case at hand, let us parametrize it
as
\begin{align}
\mathcal{S}^{\mu\nu}=\frac{2s}{|b|}\left(b^{[\mu}(v_1-\sigma v_2)^{\nu]}\right) \,.
\end{align}
Hence, after reintroducing the dimensionless velocities, we can rewrite the numerator as
\begin{align}
N_{\mathcal{S}}=-\ii s \frac{b\cdot q}{|b|}-s^2\frac{1}{2 |b|^2}(b \cdot q)^2 \;.
\end{align}
To complete the calculation we need to evaluate the following type of integrals
\begin{align}
\mathcal{I}^{\mu_1 \mu_2 \dots \mu_n}:= \int \hat \dd^{4}q \,
\hat \delta(q \cdot v_{1})\hat \delta(q \cdot v_{2})  e^{\ii q\cdot b} \frac{q^{\mu_1} \cdots q^{\mu_n}}{q^2}.
\end{align}
The simplest case $ \mathcal I^{\mu_1}$ can be computed from \eqref{I-case-simple}. For the case $\mathcal I^{\mu_1 \mu_2}$ we adapt the procedure of  Ref.~\cite{Maybee:2019jus} to our case.  Noting that the results must lie in the plane orthogonal to the four velocities the integral  $\mathcal I^{\mu_1 \mu_2}$ can be written as
\begin{align}
\mathcal I^{\mu_1 \mu_2}= c_1 b^{\mu_1} b^{\mu_2} + c_2 \Pi^{\mu_1\mu_2},
\end{align}
where $\Pi^{\mu_1\mu_2}$ is a projector which explicitly reads
\begin{align}
\Pi^{\mu_1\mu_2}=\eta^{\mu_1 \mu_2 }+v_1^{\mu_1 } v_1^{\mu_2 }-v_2^{\mu_1 } v_1^{\mu_2 }-v_1^{\mu_1 } v_2^{\mu_2 }
\end{align}
Therefore employing the traceless property of  $\mathcal I^{\mu_1 \mu_2}$ and \eqref{I-case-simple} we obtain
\begin{align}
\mathcal I^{\mu_1 \mu_2}= \frac{1}{\pi b^4  \sigma } \left(b^{\mu_1 } b^{\mu_2}-\frac{1}{2}b^{2} 
\Pi^{\mu_1\mu_2}\right).
\end{align}
After some algebra we compute
\begin{align}
\chi_1= \kappa^2\frac{p_1 \cdot p_2 }{8\pi} \left( -\frac{1}{2}\log\left(\frac{|b^2|}{L^2}\right)-\frac{s}{|b|}
+\frac{s^2 }{2|b|^2}\right),
\end{align}
which leads to the scattering angle
\begin{align}
\theta_1= 4\left(\frac{1}{|b|} - \frac{s }{|b|^2}+ \frac{s^2}{|b|^3}\right) G_N m,
\end{align}
in agreement with the  massless limit of the Kerr-result of  Ref.~\cite{Jakobsen:2021zvh} (See 
also Refs.~\cite{Ono:2017pie,Kumar:2019ohr}). 

\section{Conclusions}
\label{conclusions}
We have extended the worldline quantum field theory formalism to classical observables to the case of 
scattering of light off a massive particle.  We have constructed in a worldline representation the 
gravitationally dressed photon propagator. Then, following  the WQFT formalism, we have identified the partition function
and derived the Feynman rules to compute the eikonal phase and the deflection angle from it. 

The dressed photon propagator is built from a matrix-valued particle action whose path integral requires
time ordering. To get rid of the latter, we have introduced auxiliary variables that at the same time describe explicitly 
the spin degrees of freedom. They play an important role at the quantum level to obtain the correct amplitudes as we have
seen in Sec.\ref{photon-to-amplitude}. However, time ordering is a quantum requirement and hence the  geometric optics  regime  makes the terms related to the spin-tensor vanish, thus providing an alternative realization of the 
equivalence between the amplitude of two massive particles and that of a massive particle and a 
massless.  This simplifies the calculation  
of the eikonal phase dramatically and only two diagrams are required to calculate the NLO contribution. We have computed the NLO scattering deflection angle for spinless and LO for spinning particles, finding full agreement with the results by other methods. Our calculation is compatible with the eikonal approach based on on-shell scattering amplitudes.

An alternative approach to construct a worldline representation of the photon propagator would be to consider
the $N=2$ spinning particle, which has been used to describe the quantum theory of  spin 1 and differential forms 
in Refs.~\cite{Bastianelli:2005vk, Bastianelli:2005uy, Bastianelli:2011pe, Bastianelli:2012nh}. Indeed, in Ref.~\cite{Jakobsen:2021zvh} this model has been used to describe spinning black holes. However, the approach we have used here is closer to the standard QFT setup. It would be interesting  to compare against the massless limit of the purely worldline approach of Ref. \cite{Jakobsen:2021zvh} based on the $N=2$ spinning particle.

Our dressed propagator can also  be applied to the description of  light by light scattering, and it would be
interesting to study amplitudes made up of dressed photon propagators in the case where the spin tensor 
contributes. 
Another interesting route would be  
to explore the so-called generalized Wilson line \cite{Laenen:2008gt,White:2011yy,Bonocore:2021qxh, Bonocore:2020xuj}
for photons. 
Dressed  photon propagators are also useful in describing the propagation of light in an arbitrary medium \cite{Difallah:2019mjt} so it would be interesting to consider coupling to matter in addition to gravity. In this case, the geometric-optics regime  is not applicable and  terms associated with the spin-tensor are expected to contribute.

\addsec{Acknowledgements}
LDLC  acknowledges financial support from the {\em Open Physics Hub}
at the Physics and Astronomy Department ``Augusto Righi'' in Bologna. 

\appendix

\section{Next to leading order photon impulse}\label{app:B}
In the WQFT one can define the observables related to the point particle mechanics, by making use of Noether theorem which
in such case is nothing else but Ehrenfest theorem.
Let us first define the impulse of the photon as \cite{Mogull:2020sak}
\be 
\del p^{\mu} = \int_{-\infty}^{+\infty}\dd\sigma \, \La  \frac{\dd^{2}q^{\mu}}{\dd\sigma^{2}}\Ra = \int_{-\infty}^{+\infty} \hat\dd\omega \left(-\omega^{2}
\la 
\tilde{q}^{\mu}(\omega) \deltahat(\omega)\ra \right)\;.
\ee
The leading order calculation is straightforward and can be obtained by the evaluation of a single diagram
which, unlike  the eikonal phase, has two kinematic fluctuations, namely 
\be 
\del^{(0)}p^{\mu} = \raisebox{-10mm}{\loimp} = -\kappa^{2}\frac{\left(p_{1}\cdot p_{2}\right)^{2}}{4} 
\frac{\partial I_{\simtree}}{\partial b_\mu} =4G_{N} \left(p_{1}\cdot p_{2}\right)\frac{b^{\mu}}{|b|^{2}}\;,
\ee
where the tree level integral $I_{\simtree}$ has been evaluated in \eqref{I-case-simple}.

Let us move now to the next to leading order photon impulse.
The topologies which are vanishing for the eikonal phase do also vanish here in the same way. Particularly, the diagram with the three-graviton vertex with the two gravitons starting from the photon line is exactly zero once using momentum conservation and the delta constraints.
Thus, the only non-vanishing contributions arise from 
\ba \label{nlo-imp-one}
\raisebox{-10mm}{
	\nloimpone
}
= -\frac{\ii m^{2}\kappa^{4}}{32}\int \hat \dd^{4}q \, \deltahat(q\cdot p_{1})
\deltahat(q\cdot p_{2}) \frac{e^{\ii q\cdot  b}}{q^{2}} \int \hat \dd^{4}k \, \deltahat(k\cdot p_{2})  \frac{N_{1}(q,k,p) q^{\mu}}{k^{2} (k-q)^{2}}
\ea
with the numerator $N_{1} =
(p_1 \cdot p_2)^{2}\left(k^{2}+(k-q)^{2}\right) + m^{2}(k\cdot p_{1})^{2}$,  and the other   contributing diagram is 
\ba \label{nlo-imp-box}
\raisebox{-10mm}{
	\nloimpfour }
=\frac{\ii (p_1 \cdot p_2)^{2} \kappa^{4}}{16} \int \hat\dd^{4}q \,  \deltahat(q\cdot p_{1}) \deltahat(q\cdot p_{2}) e^{\ii q\cdot b}\\[-5mm]
\int \hat\dd^{4}k \, \deltahat(k\cdot p_{2} )\frac{N_{2}(q,k,p)\left(q-k\right)^{\mu}}{k^{2} (k-q)^{2} (k\cdot p_{1})^{2}},
\ea
where $N_{2} = (p_1 \cdot p_2)^{2}k\cdot (k-q)+2m^{2} (k\cdot p_{1})^{2}$.
Let us briefly review the integration procedure.
We first focus on Eq.\eqref{nlo-imp-one}. 
For this one we just need to evaluate the integral 
\be
{\cal I}_{1}= \int \hat\dd^{4}q \,  \deltahat(q\cdot v_{1}) \deltahat(q\cdot v_{2}) \frac{e^{\ii q\cdot b} }{q^{2}}\int \hat\dd^{4}k \, \deltahat(k\cdot v_{2} )\frac{(k\cdot v_{1})^{2}}{k^{2} (k-q)^{2}} = \frac{1}{128\pi |b|}
\ee
since we can rewrite the whole expression in the RHS of Eq.~\eqref{nlo-imp-one}  as
\be \label{imp-threeg}
-\frac{\ii E m^{2} \kappa^{4}}{32}\frac{1}{\ii}\frac{\partial }{\partial b^{\mu}} {\cal I}_{1} = -\frac{1}{4}\pi G_{N}^{2}m(p_{1}\cdot p_{2})  \frac{b^{\mu}}{|b|^{3}}\;.
\ee
Let us move now to the next diagram, which can be recast as follows
\ba \label{box}
\raisebox{-10mm}{\nloimpfour}
=\frac{\ii m^{2}k^{4} E^{2} }{16} \int \hat \dd^{4}q \, \deltahat(q\cdot v_{1})\deltahat(q\cdot v_{2})
e^{\ii q\cdot b}
\left(
2 {\cal I}_{2}^{\mu} + {\cal I}_{3}^{\mu}
\right),  
\ea
where 
\be 
{\cal I}_{2}^{\mu}= \int \hat\dd^{4}k \frac{ \deltahat(k\cdot v_{2}) (q-k)^{\mu}}{k^{2}(q-k)^{2}} ,\hskip.5cm
{\cal I}_{3}^{\mu}=\int \hat\dd^{4}k \frac{ \deltahat(k\cdot v_{2}) k\cdot (k-q) (q-k)^{\mu}}{k^{2} (q-k)^{2} (k\cdot v_{1})^{2}} \;.
\ee
Let us  perform the tensor reduction of the above integrals.
The first one can be decomposed as 
$ 
{\cal I}_{2}^{\mu} = Aq^{\mu} + B v_{2}^{\mu}
$
such that, using the delta constraint $q\cdot v_{1} = q\cdot v_{2}=0$, one finds that $B=0$. Contracting with $q^{\mu}$ one finds that 
\be 
A = \frac{1}{2} \int \hat\dd^{4}k \frac{\deltahat(k\cdot v_{2})}{k^{2} (k-q)^{2}}\;.
\ee
In this way one is able to evaluate the first contribution in \eqref{box} as 
\ba \label{second-box}
\frac{\ii m^{2}\kappa^{4}E}{8} \int \hat \dd^{4}q \, \deltahat(q\cdot v_{1})\deltahat(q\cdot v_{2})\, e^{\ii q \cdot b}{\cal I}_{2}^{\mu} = 4\pi G_{N}^{2}m (p_{1}\cdot p_{2})   \frac{b^{\mu}}{|b|^{3}}\;.
\ea
Before proceeding further, let us point out that adding up the two contributions just evaluated i.e. \eqref{imp-threeg} and \eqref{second-box} one obtains 
\be 
\frac{15\pi}{4}G_{N}^{2}m (p_{1}\cdot p_{2})
\frac{b^{\mu}}{|b|^{3}},
\ee
which  corresponds to the impulse obtained from the NLO eikonal phase, without any iteration of the scattering data.
However,  we still need to evaluate the first piece in \eqref{box}.
For this task we write 
\be 
{\cal I}_{3}^{\mu} = Aq^{\mu} + B v_{2}^{\mu} + C v_{1}^{\mu}.
\ee
Then, using that $v_{2}^{\mu}M_{\mu}=0$ one obtains $B = - \sigma C$, which allows us to rewrite 
$ 
{\cal I}_{3}^{\mu} = A q^{\mu} +C\left(v_{1}^{\mu} - \sigma v_{2}^{\mu }		\right)
$.
In this  way, after contracting with $q^{\mu}$, one finds that the integral thus obtained are tadpoles therefore concluding that $A=0$. Contracting with $v_{1}^{\mu}$ and using integral reduction allows us to fix the remaining coefficient, namely
\be
C = \frac{1}{2}q^{2} \int \hat \dd^{4}k \frac{\deltahat(k\cdot v_{2})}{
	(k^{2} +\ii\epsilon) ((k-q)^{2}+\ii\epsilon) (k\cdot v_{1}+\ii\epsilon)
} = -\frac{\ii}{2}q^{2} \int \hat\dd^{4} k \frac{\hat\delta(k\cdot v_{1})\hat\delta(k\cdot v_{2})	}{k^{2} (k-q)^{2}},
\ee
where we performed the change of variables $k-q \to -k'$ in 
the last equality and the Dirac delta representation
\be
\hat\delta(x) = \ii \left(	\frac{1}{x+\ii\epsilon}- \frac{1}{x-\ii\epsilon}\right) \;,
\ee
which enables us to rewrite the full contribution related to ${\cal I}_{3}$  as the square of  $I_{\simtree}$ which has been evaluated previously, see Eq.\eqref{I-case-simple}. 

Finally, putting all pieces together and performing the extra momentum integration, one obtains the next to leading order photon impulse
\be 
\del p^{\mu} = 4G_{N}(p_{1}\cdot p_{2}) \frac{b^{\mu}}{|b|^{2}}+G_{N}^{2}m \frac{(p_{1}\cdot p_{2})}{|b|}
\left(
\frac{15\pi  }{4} \frac{b^{\mu}}{|b|^{2}} -\frac{8 }{|b|}\left(
v_{1}^{\mu}- \sigma v_{2}^{\mu}
\right)
\right)
\ee
where we reintroduced $\sigma= v_{1}\cdot v_{2}$ to write the result in a Lorentz invariant form.

\bibliographystyle{JHEP}

 \renewcommand\bibname{References} 
\ifdefined\phantomsection		
  \phantomsection  
\else
\fi
\addcontentsline{toc}{section}{References}
\providecommand{\href}[2]{#2}\begingroup\raggedright\endgroup

\end{document}